\documentclass[%
11pt,
reprint,
onecolumn,
tightenlines,
superscriptaddress,
preprintnumbers,
nofootinbib,
amsmath,amssymb,amsthm,
physrev,
eqsecnum,
]{revtex4-2}

\usepackage{isomath}
\usepackage{amsmath,amsthm}
\usepackage{amsbsy}
\usepackage{amssymb}
\usepackage{amscd}
\usepackage{amsfonts}
\usepackage{stmaryrd}
\usepackage{siunitx}
\usepackage{euscript}
\usepackage[utf8]{inputenc}
\usepackage[T1]{fontenc}
\usepackage{newtxtext} 
\everymath{\displaystyle}
\usepackage{exscale}
\usepackage{microtype}
\usepackage{booktabs}
\usepackage{algorithm}
\usepackage{algpseudocode}
\usepackage{hyperref}

\usepackage{graphicx}
\usepackage{boxedminipage}
\usepackage{calc}
\usepackage[dvipsnames]{xcolor}
\usepackage[caption=false,justification=raggedright]{subfig}

\usepackage{setspace}
\usepackage{enumitem}
\setitemize{noitemsep,topsep=0pt,parsep=0pt,partopsep=0pt}
\setenumerate{noitemsep,topsep=0pt,parsep=0pt,partopsep=0pt}
\setdescription{noitemsep,topsep=0pt,parsep=0pt,partopsep=0pt}

\usepackage[colorinlistoftodos, color=green!40]{todonotes}
\setuptodonotes{inline,caption={}}

\usepackage{soul} % highlighting
\usepackage[normalem]{ulem}

\usepackage{orcidlink}
\usepackage{siunitx}
\usepackage[small]{titlesec}

\titlespacing*{\section}{0pt}{12pt plus 4pt minus 2pt}{2pt plus 2pt minus 2pt}
\titlespacing*{\subsection}{0pt}{12pt plus 4pt minus 2pt}{2pt plus 2pt minus 2pt}
\titlespacing*{\subsubsection}{0pt}{12pt plus 4pt minus 2pt}{2pt plus 2pt minus 2pt}
\titlespacing*{\paragraph}{0pt}{12pt plus 4pt minus 2pt}{2pt plus 2pt minus 2pt}

\makeatletter

    \renewcommand*{\p@subsection}{}
    
    \renewcommand*{\p@subsubsection}{}
\makeatother

\usepackage{upgreek}

\newcommand{\half}{\tfrac{1}{2}}

\theoremstyle{definition}

\AtEndEnvironment{definition}{\null\hfill\qedsymbol}

\AtEndEnvironment{remark}{\null\hfill\qedsymbol}

\AtEndEnvironment{example}{\null\hfill\qedsymbol}

\AtEndEnvironment{assumption}{\null\hfill\qedsymbol}

\newcommand{\dm}{\ \mathrm{d}}
\newcommand{\deriv}[2]{\frac{\dm #1}{\dm #2}}

\newcommand{\bfa}{{\mathbold a}}
\newcommand{\bfb}{{\mathbold b}}

\newcommand{\bfe}{{\mathbold e}}
\newcommand{\bff}{{\mathbold f}}

\newcommand{\bfn}{{\mathbold n}}

\newcommand{\bfC}{{\mathbold C}}

\newcommand{\bfF}{{\mathbold F}}

\newcommand{\bfI}{{\mathbold I}}

\newcommand{\bfQ}{{\mathbold Q}}
\newcommand{\bfR}{{\mathbold R}}

\newcommand{\bfU}{{\mathbold U}}

\begin{document}

\preprint{To appear in the Journal of Applied Mechanics (\href{https://doi.org/10.1115/1.4071527}{DOI:10.1115/1.4071527})}

\title{Nonlocal Linear Instability Drives the Initiation of Motion of Rational and Irrational Twin Interfaces}

\author{Chang-Tsan Lu}
    \affiliation{Department of Civil and Environmental Engineering, Carnegie Mellon University}

\author{Anthony Rollett}
    \affiliation{Department of Materials Science and Engineering, Carnegie Mellon University}

\author{Kaushik Dayal}
    \email{Kaushik.Dayal@cmu.edu}
    \affiliation{Department of Civil and Environmental Engineering, Carnegie Mellon University}
    \affiliation{Center for Nonlinear Analysis, Department of Mathematical Sciences, Carnegie Mellon University}
    \affiliation{Department of Mechanical Engineering, Carnegie Mellon University}

\date{\today}

%%%%%%%%%%%%%%%%%%%%%
%%%%%%%%%%%%%%%%%%%%%
%%%%%%%%%%%%%%%%%%%%%
%%%%%%%%%%%%%%%%%%%%%

\begin{abstract}
	Twin boundaries play a central role in the functional behavior of martensitic materials, yet the mechanisms governing the initiation of their motion remain poorly understood for twins lying along irrational crystallographic directions. 
    Here we present an atomistic investigation of the onset of motion of both rational and irrational twin interfaces in a two-dimensional model lattice with rectangular unit cells. 
    Using quasistatic shear loading and full linear stability analysis, we show that the initiation of twin boundary motion is signaled by a nonlocal linear instability, marked by the vanishing of the lowest eigenvalue of the Hessian; the corresponding eigenmode predicts the atomic displacements that initiate motion.
    We find that irrational twin boundaries have significantly lower critical shear stress to initiate motion compared to rational twin boundaries.
    Further, we find that they display unusual mechanisms to initiate motion such as the formation of microtwins in directions orthogonal to the overall twin boundary.
    Finally, we compare various local measures against the nonlocal stability analysis, and find that the former do not capture that irrational twin boundaries initiate their motion at lower stresses compared to rational boundaries.
\end{abstract}

\maketitle

%%%%%%%%%%%%%%%%%%%%%
%%%%%%%%%%%%%%%%%%%%%
%%%%%%%%%%%%%%%%%%%%%
%%%%%%%%%%%%%%%%%%%%%
\section{Introduction}\label{sec:intro}

Martensitic transformations are fundamental to the mechanical behavior of many structural and multifunctional materials, including shape-memory alloys and advanced steels \cite{christian1995deformation,bhadeshia2021theory,bhattacharya-book-2003}. 
These transformations, induced by changes in temperature or external stress, generate complex microstructures composed of multiple martensite variants separated by \emph{twin boundaries} that are coherent interfaces between variants that preserve lattice continuity.
The motion of such twin boundaries is central to the evolution of the martensitic microstructure and governs phenomena ranging from shape-memory recovery to impact response.
Consequently, understanding the motion of twin boundaries is essential for predicting and designing the response of these materials.

A defining feature of martensitic twins is their \emph{rational} or \emph{irrational} crystallographic nature, determined by the orientation of the twinning shear and normal relative to the lattice basis. 
While rational twins (e.g., type-I, type-II, and compound) have been extensively characterized, irrational or ``non-conventional'' twins \cite{pitteri-zanzotto-book-2003} --whose interfaces do not pass through rational lattice planes -- are less well understood. However, irrational twins occur in numerous transformations, including cubic-to-monoclinic and tetragonal-to-monoclinic transitions, and may exhibit distinct mechanical stability and mobility characteristics. Understanding their behavior thus offers fundamental insights into interface motion and is the focus of this work.

\paragraph*{Prior Work.}

Continuum theories have provided a framework for the evolution of twin boundary motion \cite{abeyaratne-knowles-book-2006,truskinovskii1982equilibrium,agrawal2015dynamic,agrawal2015dynamic-2,chua2022phase,chua2024interplay,yang2010formulation} but these models require essential input from the atomic-scale. 
Consequently, there has been significant work on understanding atomic scale mechanisms.
Using molecular dynamics, \cite{abeyaratne-vedantam-2003} demonstrated that compound twins exhibit higher mobility than type-I twins; \cite{Sinha_Kulkarni_2014} showed that rational twin boundaries can be intrinsically brittle in one crystallographic direction and ductile in the opposite direction, explaining anisotropic fracture responses in nanotwinned metals;  \cite{Hammami_Kulkarni_2014} studied nanopillar compression to reveal how twin spacing and specimen size interact to control yield and post-yield behavior, exposing distinct size-dependent deformation modes; \cite{Jiao_Kulkarni_2015} identified twin-boundary migration as a controlling creep mechanism at intermediate/high stresses; and \cite{Chen_Kulkarni_2013} demonstrate that rational twin boundaries exhibit unusual fluctuation kinetics and long-range interactions.
More recent studies have examined the role of interface structure and energetics \cite{mohammed2020martensitic,mohammed2020modeling,gengor2021101,hunter2014predictions,hunter2015relationship} on twin nucleation and related properties.
On the experimental front, \cite{vollach2017kinetics,faran2011kinetic,faran2013kinetic,faran2010twin} measured twin kinetics in NiMnGa but these are challenging experiments to conduct and interpret.

Despite these advances, the initiation of twin boundary motion, particularly for irrational twins, has received little attention. 
In our prior work \cite{lu-dayal-2011}, we demonstrated that, in rational twins, the onset of motion corresponds to a loss of linear stability, where the lowest eigenvalue of the system Hessian matrix approaches zero and the associated eigenmode predicts the initial displacement pattern. 
However, how this nonlocal instability mechanism generalizes to irrational twin interfaces remains unclear.
Nonlocality has also been highlighted in related contexts such as dislocation nucleation~\cite{miller2008nonlocal}.

\paragraph*{Contributions of This Work.}

In this work, we use a model atomistic framework, based on \cite{hildebrand-abeyaratne-2008}, to study the initiation of motion in irrational twin boundaries and compare it with their rational counterparts. 
We construct atomic configurations containing twin interfaces with both rational and irrational orientations. 
Under quasistatic shear loading and energy minimization, we perform linear stability analyses of the Hessian.
We find that the initiation of twin boundary motion is governed by a \emph{nonlocal linear instability}, manifested as the vanishing of the lowest Hessian eigenvalue. 
The corresponding eigenmodes accurately predict the initial displacement fields leading to detwinning. 
Our results further reveal that irrational twins exhibit lower critical stresses and distinct instability patterns compared to rational twins, indicating that their structure enhances mobility. 
Specifically, we find that irrational twins exhibit microtwin formation in the twin boundary motion process, complementing other mechanisms such as ledge motion \cite{hildebrand-abeyaratne-2008,zhen2008dynamics,zhen2008dynamics}.
These findings provide a mechanistic picture that extends the understanding of interface motion to the broader class of irrational twins.
We also compare a range of local criteria to examine whether they can identify differences in the critical shear stress for the initiation of twin boundary motion.
We find that these are not able to show any correlations with the behavior observed in molecular dynamics, while the nonlocal linear instability provides a very good indicator, suggesting the inherent nonlocality of detwinning, in analogy with the findings in other defect systems \cite{miller2008nonlocal}.

%%%%%%%%%%%%%%%%%%%%%%%%%%%%%%%%%%%%%%%%%%%%%%%%%%
%%%%%%%%%%%%%%%%%%%%%%%%%%%%%%%%%%%%%%%%%%%%%%%%%%
%%%%%%%%%%%%%%%%%%%%%%%%%%%%%%%%%%%%%%%%%%%%%%%%%%
%%%%%%%%%%%%%%%%%%%%%%%%%%%%%%%%%%%%%%%%%%%%%%%%%%
\section{Classical Continuum Description of Twinning}\label{sec:twin-continuum}

%%%%%%%%%%%%%%%%%%%%%%%%%%%%%%%%%%%%%%%%%%%%%%%%%
\begin{figure}[htb!]
  \centering
    \includegraphics[width=0.9\textwidth]{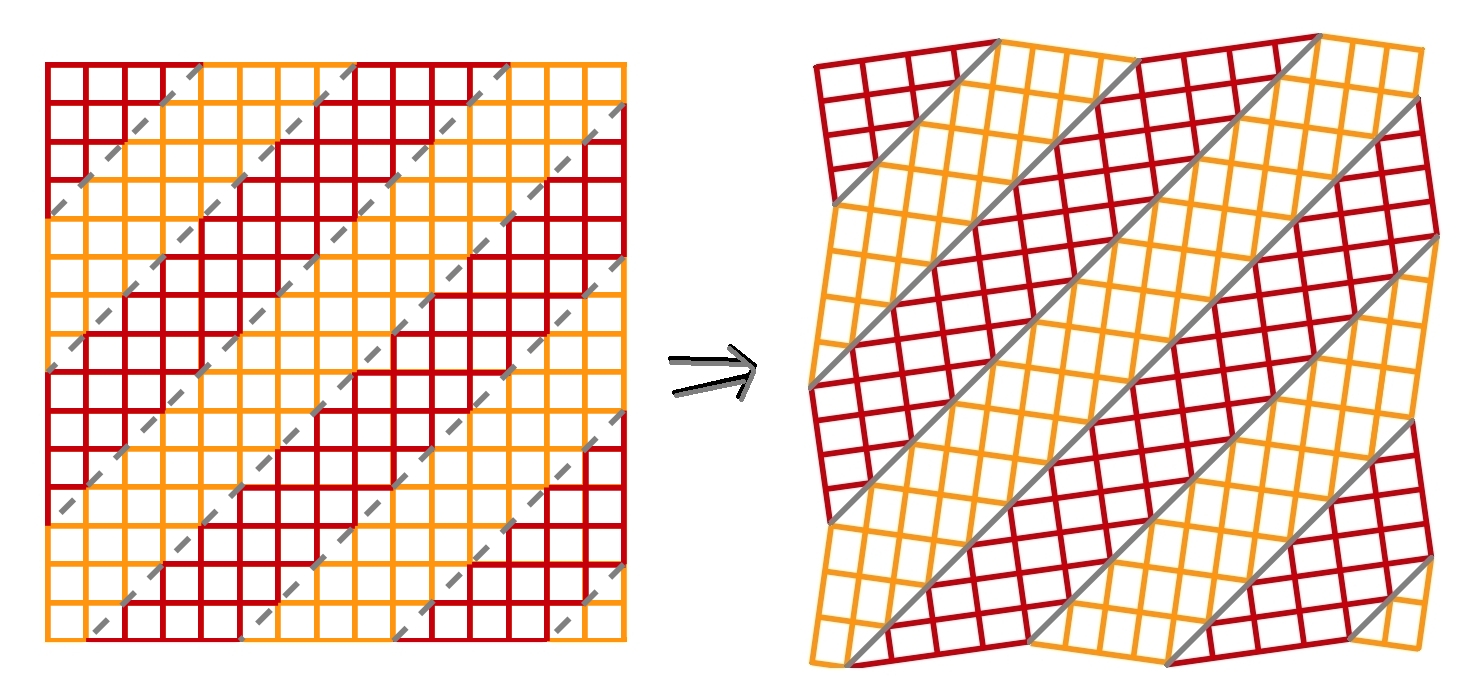}
  \caption{The square austenitic lattice (left) is transformed into a twinned rectangular martensitic lattice (right), with the coherent twin bounmdaries shown in gray.}
  \label{fig:austenite-to-twinned_martensite}
\end{figure}

%%%%%%%%%%%%%%%%%%%%%%%%%%%%%%%%%%%%%%%%%%%%%%%%%

We briefly summarize here the relevant aspects of the continuum theory of austenite-martensite phase transformations and twinning (Fig. \ref{fig:austenite-to-twinned_martensite}).
The definition of a twin boundary is \cite{christian1995deformation} given by the 2 conditions:
\begin{enumerate}
    \item The lattice on one side can be obtained by a simple shear of the lattice on the other;
    \item The lattice on one side can also be obtained by a rotation of the lattice on the other.
\end{enumerate}

When considering the deformations occurring in an austenite-martensite transformation, it is convenient to take the austenite phase as the reference configuration.
Suppose that there are $N$ possible martensite variants, and the deformations relative to the reference configuration for these $N$ variants can be described by the corresponding deformation gradients, $\bfF_1, \bfF_2, \ldots, \bfF_N$.
Each of the deformation gradients can be uniquely decomposed as $\bfF_{I} = \bfQ_{I} \bfU_{I}, I=1\ldots N$ through the polar decomposition, where $\bfQ_{I}$ are rotations and $\bfU_{I}$ are the symmetric and positive-definite Bain tensors.

We denote by $\bfR$ the elements of the point group, i.e., the set of rotations that maps the austenite lattice back to itself.
Then, we have that there exists a $\bfR$ that relates any pair of martensitic variants by:
\begin{equation}\label{eq:variant-relation}
     \bfU_{J} = \bfR^{T}\bfU_{I}\bfR
\end{equation}

\begin{figure}[htb!]
    \centering
    \includegraphics[width=0.75\textwidth]{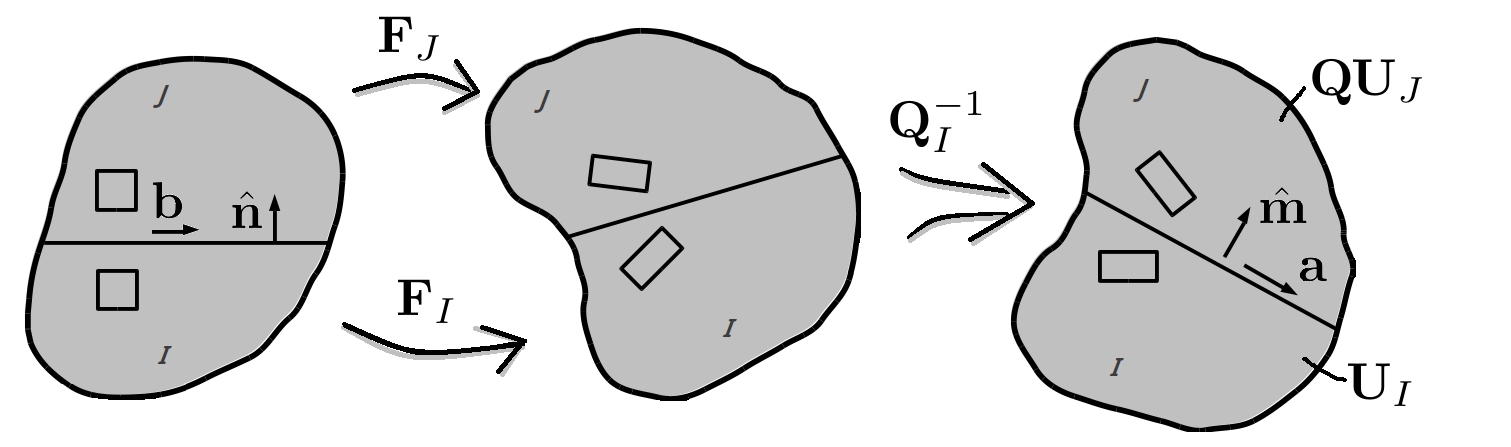}
    \caption{Left: reference configuration (austenite). Center: the region below the twin interface is transformed through $\bfF_{I}=\bfQ_{I}\bfU_{I}$ and the region above is transformed through $\bfF_{J}=\bfQ_{J}\bfU_{J}$. Right: an overall rotation of $\bfQ_{I}^{-1}$ is applied.}
    \label{fig:kinematic-compatibility}
\end{figure}

We consider a twin interface as in Figure \ref{fig:kinematic-compatibility}.
The deformation across the interface must satisfy the kinematic compatibility condition:
\begin{equation}\label{eq:kinematic-compatibility}
    \bfF_J - \bfF_I = \bfb \otimes \hat\bfn
    \implies
    \bfQ_J\bfU_{J}-\bfQ_I\bfU_{I} = \bfb\otimes\hat{\bfn}
    \implies
    \bfQ \bfU_J - \bfU_I = \bfa\otimes\hat{\bfn}
\end{equation}
where $\bfQ := \bfQ_I^T \bfQ_J$; $\hat{\bfn}$ is the unit normal of the twin interface in the reference configuration; and $\bfa := \bfQ_I^T \bfb$ is the twinning shear in the deformed configuration.

It has been shown \cite{ball-james-1987} that if $\bfC := \bfU^{-1}_{I}\bfU^{2}_{J}\bfU^{-1}_{I} \neq \bfI$ has eigenvalues that satisfy:
\begin{equation}
    \lambda_{1}\leq 1, \quad \lambda_{2}=1, \quad \lambda_{3}\geq 1
\end{equation}
then there are two sets of solutions for \eqref{eq:kinematic-compatibility}:
\begin{equation}\label{eq:ball-james-solution}
     \begin{split}
    \bfa & = \rho\left(\sqrt{\frac{\lambda_{3}\left(1-\lambda_{1}\right)}{\lambda_{3}-\lambda_{1}}}\hat{\bfe}_{1}\pm\sqrt{\frac{\lambda_{1}\left(\lambda_{3}-1\right)}{\lambda_{3}-\lambda_{1}}}\hat{\bfe}_{3}\right),
    \\
    \hat{\bfn} & = \frac{\sqrt{\lambda_{3}}-\sqrt{\lambda_{1}}}{\rho\sqrt{\lambda_{3}-\lambda_{1}}}\left(-\sqrt{1-\lambda_{1}}\bfU_{I}\hat{\bfe}_{1}\pm\sqrt{\lambda_{3}-1}\bfU_{I}\hat{\bfe}_{3}\right),
    \\
    \bfQ & = \left(\bfa\otimes\hat{\bfn}+\bfU_{I}\right)\bfU_{J}.
    \end{split}
\end{equation}
where $\hat{\bfe}_{1}$ and $\hat{\bfe}_{3}$ are the eigenvectors of $\bfC$ corresponding to $\lambda_{1}$ and $\lambda_{3}$, and $\rho$ is determined such that $\hat{\bfn}$ is normalized; the choices of $\pm$ must be made consistently.

Twins can be classified into rational and irrational boundaries based on the character of $\bfa$ and $\hat{\bfn}$.
In principle, a crystal direction is rational if, when written in the lattice basis, the ratios of the components can be written as rational numbers; in practice, the rational numbers should have small integers for the numerators and the denominators.
If the rational representation involves large integers, it is said to be an irrational direction.
It follows that a rational direction passes through a closely-spaced set of lattice points.
If $\hat{\bfn}$ is rational but $\bfa$ is irrational, the twin is called \textit{type-I}; if $\bfa$ is rational but $\hat{\bfn}$ is irrational, it is called \textit{type-II}; if both are rational, it is a \textit{compound} twin.
If both $\bfa$ and $\hat{\bfn}$ are irrational, it is sometimes called a \textit{non-conventional} twin boundary \cite{pitteri-zanzotto-1998}.

In physical terms, irrational lattice directions imply that the interfacial atoms see a wide variety of bonding environments.
On the other hand, rational interfaces have a much more limited variety of bonding environments.

%%%%%%%%%%%%%%%%%%%%%%%%%%%%%%%%%%%%%%%%%%%%%%%%%

%%%%%%%%%%%%%%%%%%%%%%%%%%%%%%%%%%%%%%%%%%%%%%%%%
%%%%%%%%%%%%%%%%%%%%%%%%%%%%%%%%%%%%%%%%%%%%%%%%%%
%%%%%%%%%%%%%%%%%%%%%%%%%%%%%%%%%%%%%%%%%%%%%%%%%%
%%%%%%%%%%%%%%%%%%%%%%%%%%%%%%%%%%%%%%%%%%%%%%%%%%
%%%%%%%%%%%%%%%%%%%%%%%%%%%%%%%%%%%%%%%%%%%%%%%%%%
\section{Construction of Irrational Twin Interfaces}
\label{sec:irrational-lattice}

For simplicity, we aim to work in two dimensions.
The simplest structural transformation would then be a square-rectangle transition.
However, the point group for this transformation is small, i.e., there are not many choices for $\bfR$ in \eqref{eq:variant-relation}.
This implies that there is a limited set of solutions to \eqref{eq:kinematic-compatibility}, and in fact all the solutions are compound twins that are symmetry-related.

To be able to compare a large number of qualitatively-different interfaces, we consider an austenitic point group that is composed of all rotations, providing us the most freedom in obtaining twinning interface solutions from \eqref{eq:variant-relation}, \eqref{eq:kinematic-compatibility}.
While an isotropic crystalline phase is strictly not realizable physically, the martensitic twin interfaces that are obtained from this procedure are coherent, physical and realizable, and satisfy the definition of twins as lattices that are related by both a shear and a rotation.

Considering the deformation shown in Figure \ref{fig:kinematic-compatibility}, and let $\left\lbrace \bfe^{\left(sq\right)}\right\rbrace$, $\left\lbrace \bfe^{\left(I\right)}\right\rbrace$ and $\left\lbrace \bfe^{\left(J\right)}\right\rbrace$ be, respectively, the lattice vectors of the parent square lattice and the lattices on the two sides of the interface after deformation.
The deformed lattice vectors can be determined by the action of the deformation gradient on the referential lattice vectors, giving:
\begin{equation}\label{eq:cauchy-born}
    \bfe_{i}^{\left(I\right)}=\bfU_{I}\bfe_{i}^{\left(sq\right)} \text{ and } \bfe_{i}^{\left(J\right)}=\bfQ\bfU_{J}\bfe_{i}^{\left(sq\right)}.
\end{equation}
where $i = 1, 2$.

We consider the first requirement of twinning, that the lattices on either side of the twin interface be related by a shear.
From \eqref{eq:kinematic-compatibility}, we have
\begin{equation}\label{eq:kinematic-compatibility-2D}
    \bfQ\bfU_{J}=\bfU_{I}+\bfa\otimes\hat{\bfn}=\left[\bfI+\bfa\otimes\left(\bfU_{I}^{-1}\hat{\bfn}\right)\right]\bfU_{I}.
\end{equation}
From \eqref{eq:variant-relation}, we have $\det \bfU_{I} = \det \bfU_{J}$.
Taking the determinant of both sides of \eqref{eq:kinematic-compatibility-2D}, we find:
\begin{equation}
    \det\bfU_{J}=\det\left(\bfQ\bfU_{J}\right)=\det\left[\bfI+\bfa\otimes\left(\bfU_{I}^{-1}\hat{\bfn}\right)\right]\det\bfU_{I} 
    \implies 
    \det\left[\bfI+\bfa\otimes\left(\bfU_{I}^{-1}\hat{\bfn}\right)\right]=1
\end{equation}
Consequently, $\bfa$ is orthogonal to $\bfU_{I}^{-1}\hat{\bfn}$, and $\bfI+\bfa\otimes\left(\bfU_{I}^{-1}\hat{\bfn}\right)$ is a simple shear.

Applying both sides of \eqref{eq:kinematic-compatibility-2D} to $\left\lbrace \bfe^{\left(sq\right)}\right\rbrace$ and using \eqref{eq:cauchy-born} gives:
\begin{equation}
    \bfe_{i}^{\left(J\right)}=\left[\bfI+\bfa\otimes\left(\bfU_{I}^{-1}\hat{\bfn}\right)\right]\bfe_{i}^{\left(I\right)}.
\end{equation}
Hence, the lattices on either side of the twin interface are related by a shear.

We consider the second requirement of twinning, that the lattices on either side of the twin interface be related by a rotation.
We use \eqref{eq:variant-relation} and \eqref{eq:cauchy-born}:
\begin{equation}
    \bfe_{i}^{\left(J\right)} 
    =
    \bfQ\bfU_{J}\bfe_{i}^{\left(sq\right)}
    =
    \bfQ\bfR^{T}\bfU_{I}\bfR\bfe_{i}^{\left(sq\right)}
    =
    \bfQ\bfR^{T}\bfU_{I}\bfe_{i}^{\left(sq\right)}\bfR^{T}
    =
    \bfQ\bfR^{T}\bfe_{i}^{\left(I\right)}\bfR^{T}
    =
    \bfQ\bfR^{T}\bfR\bfe_{i}^{\left(I\right)}
    =
    \bfQ\bfe_{i}^{\left(I\right)}
\end{equation}
where we use $\bfR^T \bfR = \bfI$.
This shows that the rotation $\bfQ$ acts on $\left\lbrace\bfe^{\left(I\right)}\right\rbrace$ to give $\left\lbrace\bfe^{\left(J\right)}\right\rbrace$.
Hence, the lattices on either side of the twin interface are related by a rotation.

To construct the interfaces explicitly, we represent $\bfR$ through the arbitrary angle $\theta_{1}$ with axis out-of-plane:
\begin{equation}
    \bfR = \left[\begin{array}{cc} \cos\theta_{1} & \sin\theta_{1} \\ -\sin\theta_{1} & \cos\theta_{1}\end{array}\right]
\end{equation}
Since the martensite is set be rectangular, from \eqref{eq:variant-relation} we have:
\begin{equation}
    \bfU_{I} = \left[\begin{array}{cc} \mu & 0 \\ 0 & \nu \end{array}\right]
    \quad \text{ and } \quad
    \bfU_{J} 
    = 
    \left[\begin{array}{cc} \cos\theta_{1} & \sin\theta_{1} \\ -\sin\theta_{1} & \cos\theta_{1}\end{array}\right]
    \left[\begin{array}{cc} \mu & 0 \\ 0 & \nu \end{array}\right]
    \left[\begin{array}{cc} \cos\theta_{1} & -\sin\theta_{1} \\ \sin\theta_{1} & \cos\theta_{1}\end{array}\right]
\end{equation}
where $\mu$ and $\nu$ are specified.

The twinning equation \eqref{eq:kinematic-compatibility} can then be explicitly written out in terms of components as:
\begin{equation}
\label{eq:system-nonlinear-equation}
\begin{split}
  \left[\begin{array}{cc} \cos\theta_{2} & -\sin\theta_{2} \\ \sin\theta_{2} & \cos\theta_{2}\end{array}\right]
  \left[\begin{array}{cc} \cos\theta_{1} & \sin\theta_{1} \\ -\sin\theta_{1} & \cos\theta_{1}\end{array}\right]
  \left[\begin{array}{cc} \mu & 0 \\ 0 & \nu \end{array}\right]
  \left[\begin{array}{cc} \cos\theta_{1} & -\sin\theta_{1} \\ \sin\theta_{1} & \cos\theta_{1}\end{array}\right]
  \\
  -
  \left[\begin{array}{cc} \mu & 0 \\ 0 & \nu \end{array}\right]
  = \left[\begin{array}{cc} a_{1}\hat{n}_{1} & a_{1}\hat{n}_{2} \\ a_{2}\hat{n}_{1} & a_{2}\hat{n}_{2} \end{array}\right],
\end{split}
\end{equation}
where $a_{i}$ and $\hat{n}_{i}$ are components with respect to a set of orthonormal basis.

\begin{figure}[htb!]
    \centering
    \subfloat[{$\bfa=\left[130\right]_{1}$}]
        {\label{fig:nx01ny03s01}\includegraphics[width=0.48\textwidth]{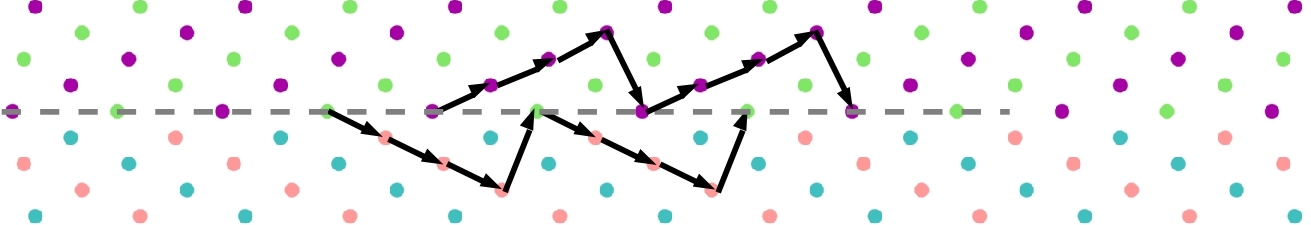}}
    \hfill
    \subfloat[{$\bfa=\left[130\right]_{2}$}]
        {\label{fig:nx01ny03s02}\includegraphics[width=0.48\textwidth]{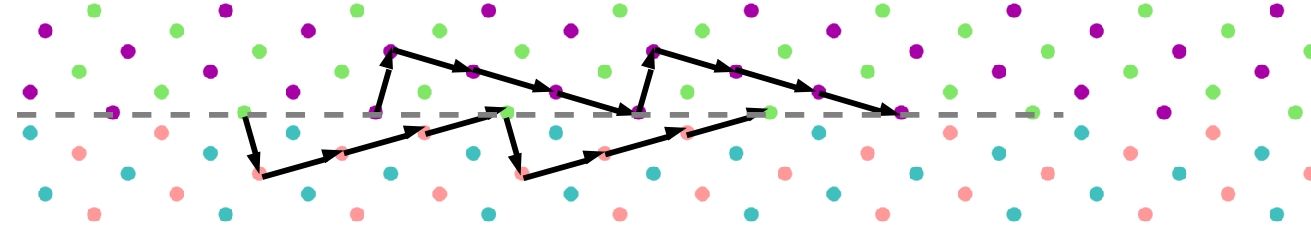}}
    \\
    \subfloat[{$\bfa=\left[190\right]_{1}$}]
        {\label{fig:nx01ny09s01a}\includegraphics[width=0.48\textwidth]{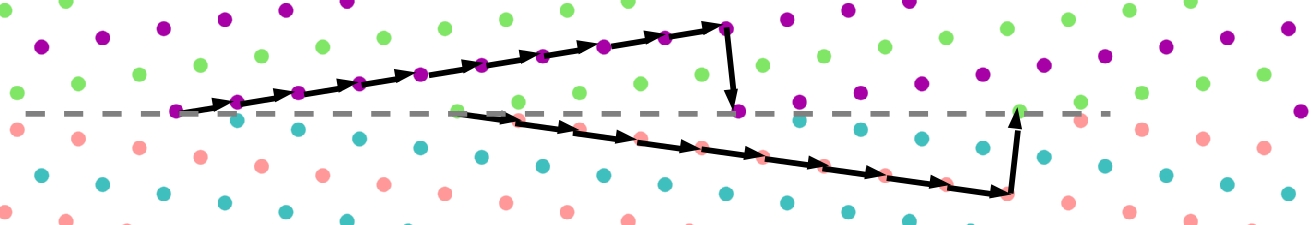}}
    \hfill
    \subfloat[{$\bfa=\left[190\right]_{2}$}]
        {\label{fig:nx01ny09s02a}\includegraphics[width=0.48\textwidth]{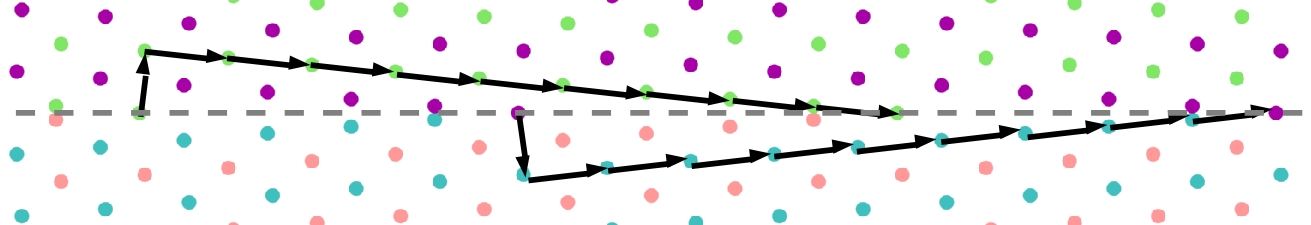}}
    \\
    \subfloat[{$\bfa=\left[250\right]_{1}$}]
        {\label{fig:nx02ny05s01a}\includegraphics[width=0.48\textwidth]{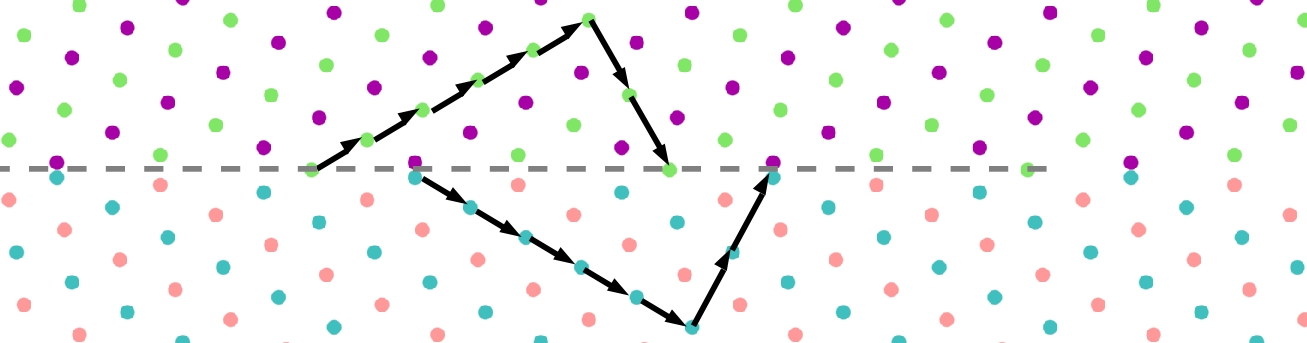}}
    \hfill
    \subfloat[{$\bfa=\left[250\right]_{2}$}]
        {\label{fig:nx02ny05s02a}\includegraphics[width=0.48\textwidth]{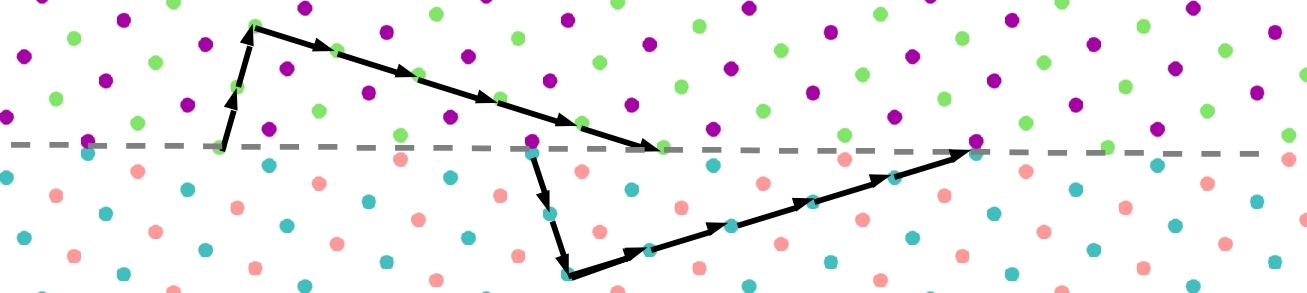}}
    \caption{Some examples of the structure of twin boundaries, constructed using \eqref{eq:system-nonlinear-equation}. Gray lines mark the twin boundaries, and green and pink dots denote the different atomic species. 
    The arrows show the periodicity along the interface: for $\bfa=\left[ij0\right]_1$, we traverse $i \bff_1 + j \bff_2$ to go from one interface atom to the next; and for $\bfa=\left[ij0\right]_2$, we traverse $i \bff_2 + j \bff_1$.}
  \label{fig:irrational-twin-plane}
\end{figure}

We can specify one of the six quantities $\theta_{1}$, $\theta_{2}$, $a_{1}$, $a_{2}$, $\hat{n}_{1}$ and $\hat{n}_{2}$ -- noting that $\hat{n}_{1}$ and $\hat{n}_{2}$ are related by the unit vector constraint -- and solve the four nonlinear equations in \eqref{eq:system-nonlinear-equation} for the unspecified quantities.
To obtain a diverse variety of rational and irrational interfaces, we specify $\theta_{1}$ and solve for the other quantities; once we obtain them, we rotate the overall system to align the twin interface normal with the vertical direction to ensure compatibility with periodic boundary conditions.
We denote the martensitic lattice vectors in this rotated simulation frame by $\bff_1$ and $\bff_2$.
Figure \ref{fig:irrational-twin-plane} gives some examples of twin boundaries.

%%%%%%%%%%%%%%%%%%%%%%%%%%%%%%%%%%%%%%%%%%%%%%%%%

%%%%%%%%%%%%%%%%%%%%%%%%%%%%%%%%%%%%%%%%%%%%%%%%%

\section{Simulation Methods}\label{sec:simulation}

%%%%%%%%%%%%%%%%%%%%%%%%%%%%%%%%%%%%%%%%%%%%%%%%%
\begin{figure}[htb!]
  \centering
    \includegraphics[width=0.75\textwidth]{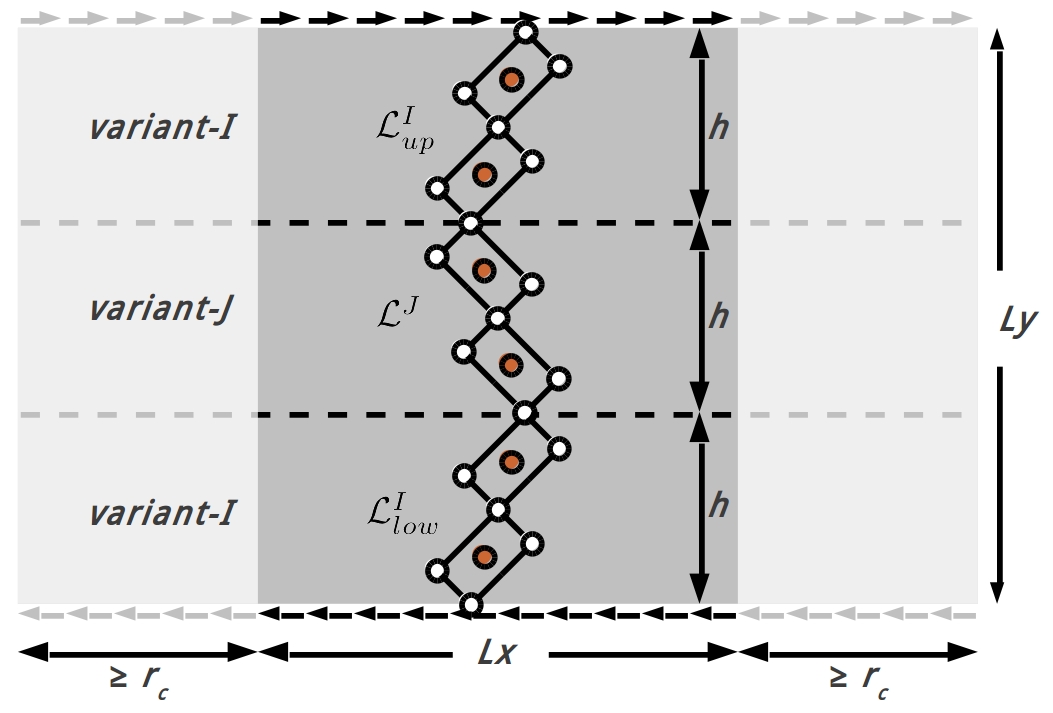}
  \caption{Schematic of the periodic simulation cell. The twin planes are aligned horizontally and the box is sheared as indicated by the arrows.}
  \label{fig:simulation-box}
\end{figure}
%%%%%%%%%%%%%%%%%%%%%%%%%%%%%%%%%%%%%%%%%%%%%%%%%

We construct the atomic configuration as schematically shown in Figure \ref{fig:simulation-box} to be compatible with periodic boundary conditions.
While strictly irrational twin planes would require an infinite periodicity in the horizontal direction, we use a large but finite periodicity to mimic this.
We require that the normal unit vector of the twin boundary is in the vertical direction to be compatible with periodicity and we ensure this by imposing the appropriate overall rotation.
We will use a model biatomic system, and, by symmetry, the second species will be located at the center of the unit cells.

We follow \cite{hildebrand-abeyaratne-2008} in setting up the atomic interactions.
\cite{hildebrand-abeyaratne-2008} developed a model biatomic system that qualitatively mimicked a two-dimensional model of Ni-Mn shape memory alloy that has rectangular unit cells.
They describe the interaction by Lennard-Jones pair potentials; for any two atoms $i$ and $j$,
\begin{equation}
    u_{\alpha\beta}(r^{ij}) = 4\epsilon_{\alpha\beta}\left(\left(\frac{\sigma_{\alpha\beta}}{r^{ij}}\right)^{12}-\left(\frac{\sigma_{\alpha\beta}}{r^{ij}}\right)^{6}\right).
\end{equation}
where $\alpha$ and $\beta$ denote the species of the atoms, either Ni or Mn here, and $r^{ij}$ is the distance between atoms.
The parameters are $\sigma_{\alpha\beta} = \{1.0, 1.0205, 0.8736\}$ and $\epsilon_{\alpha\beta}=\{1.0, 0.9810, 0.9905\}$ for Ni-Ni, Ni-Mn, and Mn-Mn respectively, where we have nondimensionalized by setting $\sigma_{NiNi}$ and $\epsilon_{NiNi}$ as the units for length and energy.

Because we aim to examine the linearized modes, we require the potential to be sufficiently differentiable.
Therefore, we modify the potential by shifting to provide a smooth cut-off at $r^{c}=4.5\sigma_{NiNi}$ both in energy and force:
\begin{equation}
    \phi_{\alpha\beta}(r^{ij}) = u_{\alpha\beta}(r^{ij})-\left(r^{ij}-r^{c}\right)\left.\deriv{u_{\alpha\beta}}{r^{ij}}\right|_{r^{c}}-u_{\alpha\beta}(r^{c}).
\end{equation}
The lattice parameters of the rectangular ground state displayed by this potential are of $\mu=1.13999$ and $\nu=1.54207$; these are slightly different from those given by Hildebrand and Abeyaratne \cite{hildebrand-abeyaratne-2008} because the shifting and smoothing of the potential.

%%%%%%%%%%%%%%%%%%%%%%%%%%%%%%%%%%%%%%%%%%%%%%%%%
   
%%%%%%%%%%%%%%%%%%%%%%%%%%%%%%%%%%%%%%%%%%%%%%%%%

Section \ref{sec:irrational-lattice} gives us the idealized structure of the twin interface based on the approximations of continuum mechanics.
We begin our simulations by setting up the idealized structure, and then relaxing the structure until it achieves equilibrium.
Continuum twinning theory gives a very good approximation to the relaxed interface structures and the atomic displacements to relax are very small.

Once we have the equilibrium structure, we then apply a shear loading using the standard Parinello-Rahman method \cite{parrinello1981polymorphic}; all our calculations are at zero temperature.
Once we reach equilibrium under load -- defined by requiring the net force on every atom to be less than $10^{-6}$ in the nondimensionalized scale -- we compute the Hessian matrix and examine the eigenvalues and eigenvectors to identify the soft modes (if any).
We then increment the load and repeat the process.

%%%%%%%%%%%%%%%%%%%%%%%%%%%%%%%%%%%%%%%%%%%%%%%%%

\section{Results and Discussion}\label{sec:results-discussion}

\begin{figure}[htb!]
    \centering
    \subfloat[The evolution of the 10 lowest eigenvalues of Hessian matrix.]
        {\label{fig:nx01ny06s01a01eigenvalues}\includegraphics[width=0.4\textwidth]{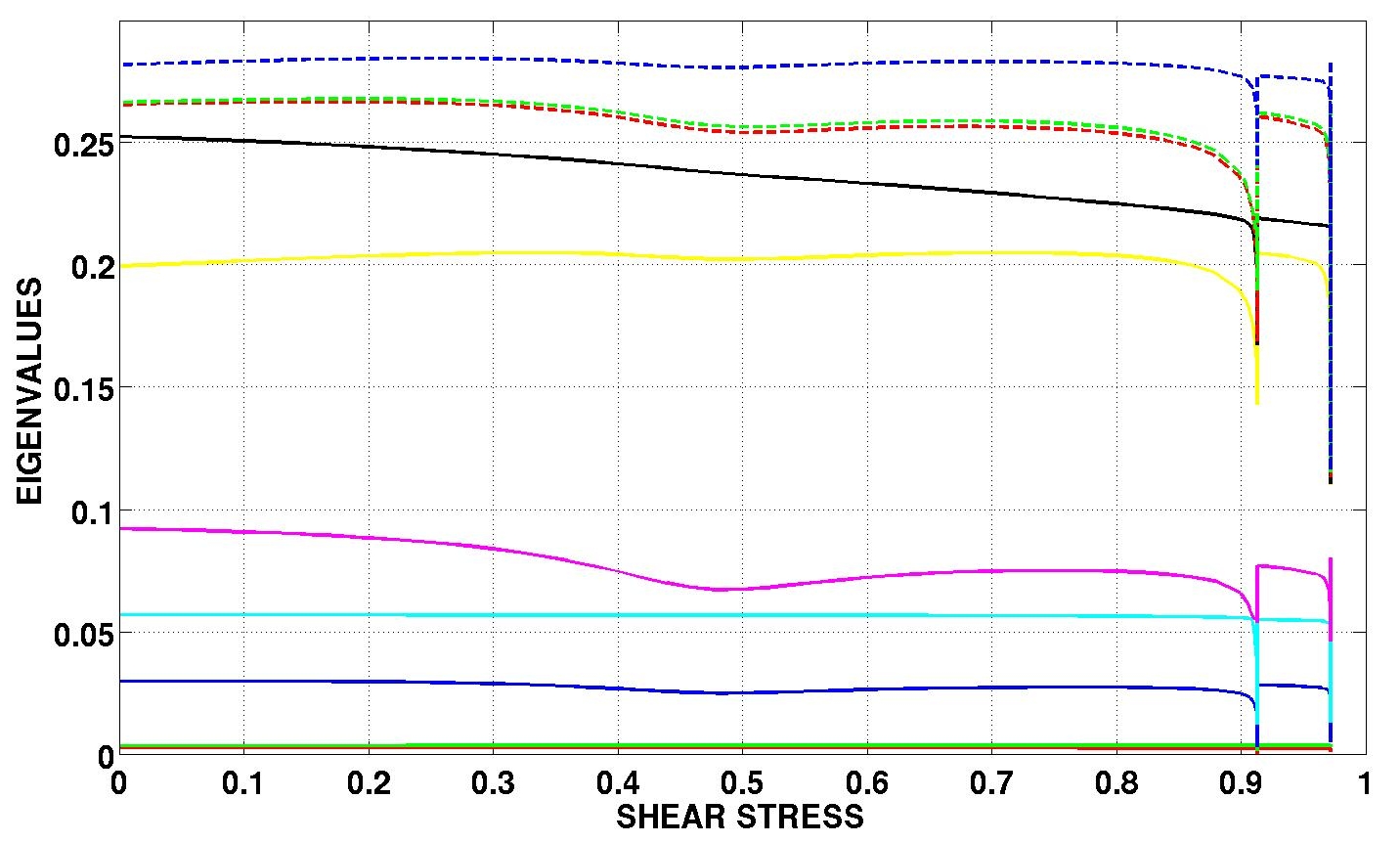}}
    \subfloat[After the first drop.]
        {\label{fig:nx01ny06s01a01current069}\includegraphics[width=0.3\textwidth]{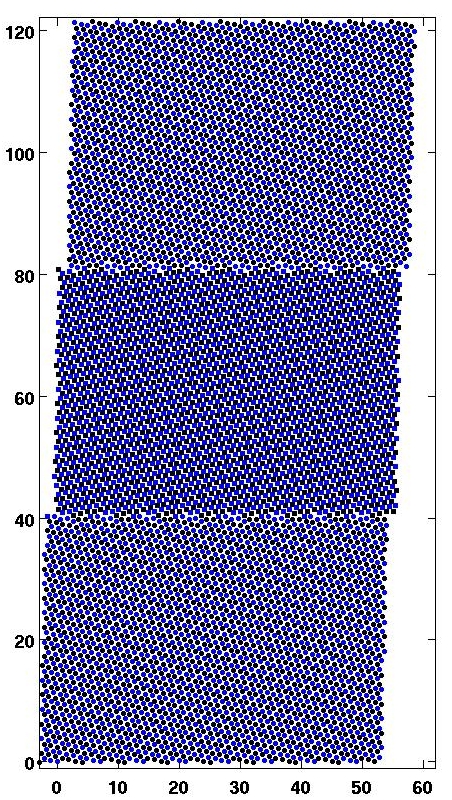}}    
    \hfill
    \subfloat[After the second drop.]
        {\label{fig:nx01ny06s01a01current091}\includegraphics[width=0.3\textwidth]{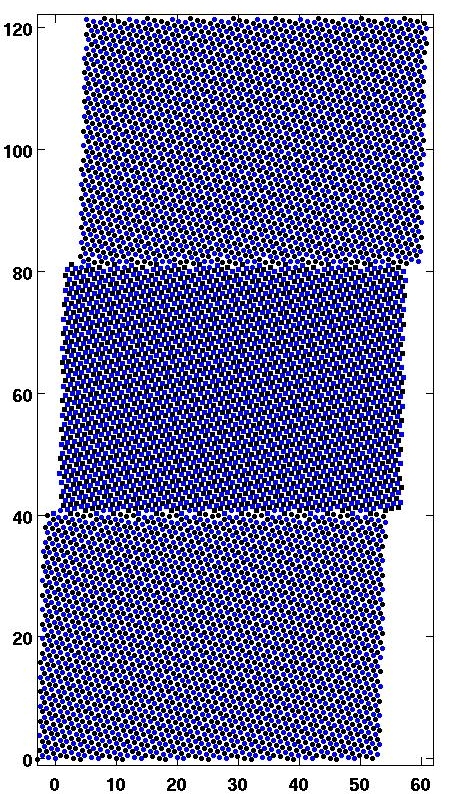}}  
    \caption{Twin boundary motion for $\bfa=\left[160\right]_{1}$: (a) shows the evolution of the 10 lowest eigenvalues with load; (b) and (c) show the atomic configuration after the first and second load drops.}
    \label{fig:nx01ny06s01a01}
\end{figure}

\begin{figure}[htb!]
    \centering
    \subfloat[Eigenmode around the upper twin boundary corresponding to Fig. \ref{fig:nx01ny06s01a01current069}.]
        {\label{fig:nx01ny06s01a01mode068up}\includegraphics[width=0.47\textwidth]{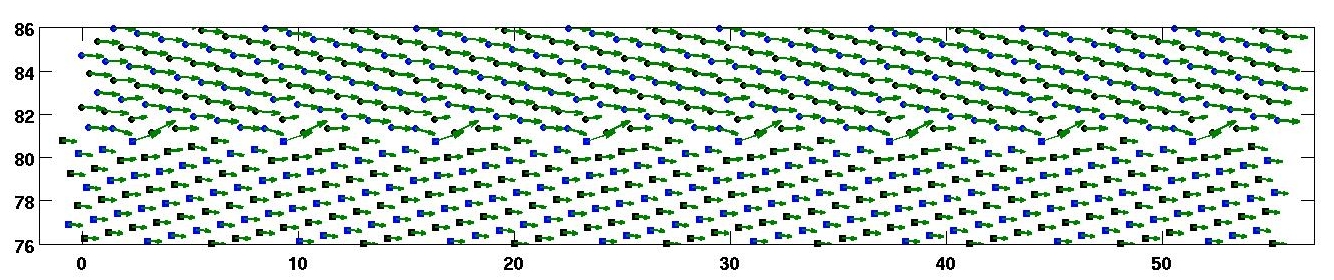}}
    \hfill
    \subfloat[Displacement around the upper twin boundary of Fig. \ref{fig:nx01ny06s01a01current069}.]
        {\label{fig:nx01ny06s01a01disp068up}\includegraphics[width=0.47\textwidth]{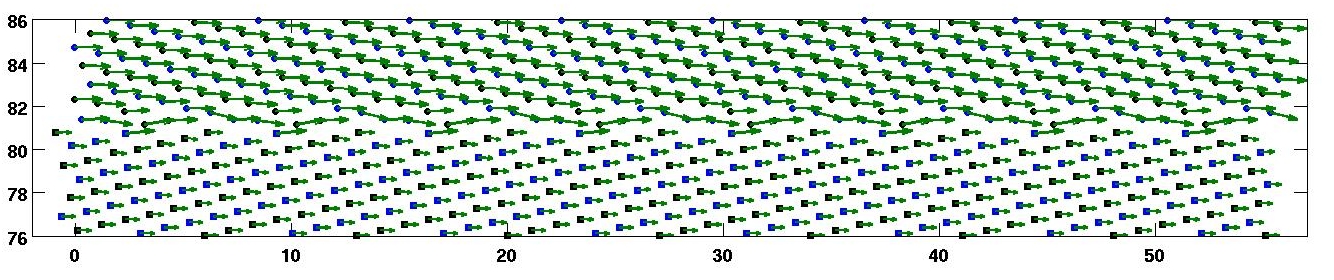}}
    \\
    \subfloat[Eigenmode around the lower twin boundary corresponding to Fig. \ref{fig:nx01ny06s01a01current069}.]
        {\label{fig:nx01ny06s01a01mode068low}\includegraphics[width=0.48\textwidth]{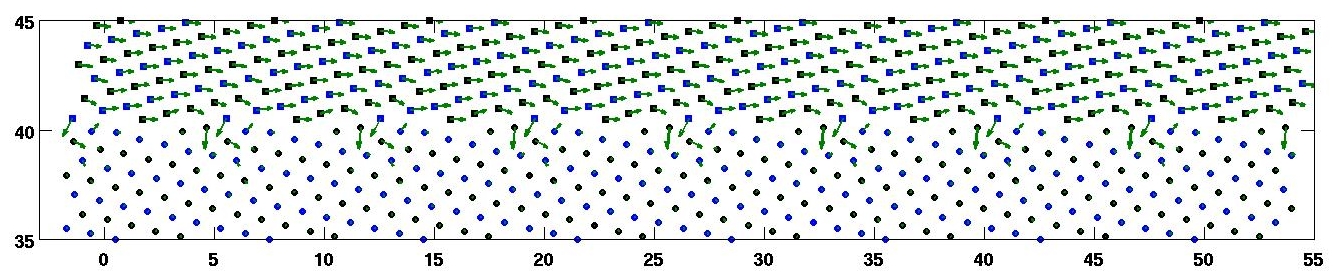}}
    \hfill
    \subfloat[Displacement around the lower twin boundary of Fig. \ref{fig:nx01ny06s01a01current069}.]
        {\label{fig:nx01ny06s01a01disp068low}\includegraphics[width=0.48\textwidth]{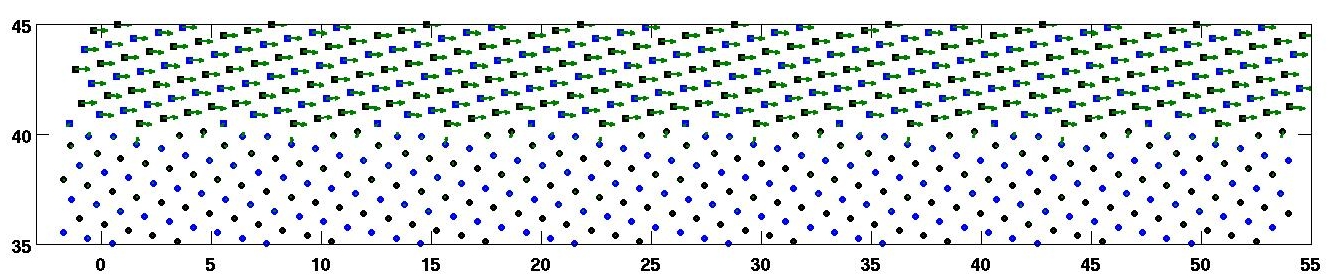}}
    \caption{Comparison of the soft eigenmodes and the observed displacements for $\bfa=\left[160\right]_{1}$ at the first eigenvalue drop, zoomed in to the vicinity of the twin interfaces. We see that the eigenmodes predict the displacements accurately. The eigenmodes and real displacements corresponding to the second eigenvalue drop are similar to the first drop and are not shown here.}
    \label{fig:nx01ny06s01a01-eigenmode}
\end{figure}

\begin{figure}[htb!]
    \centering
    \subfloat[The evolution of the 10 lowest eigenvalues of Hessian matrix.]
        {\label{fig:nx01ny06s02m03eigenvalues}\includegraphics[width=0.4\textwidth]{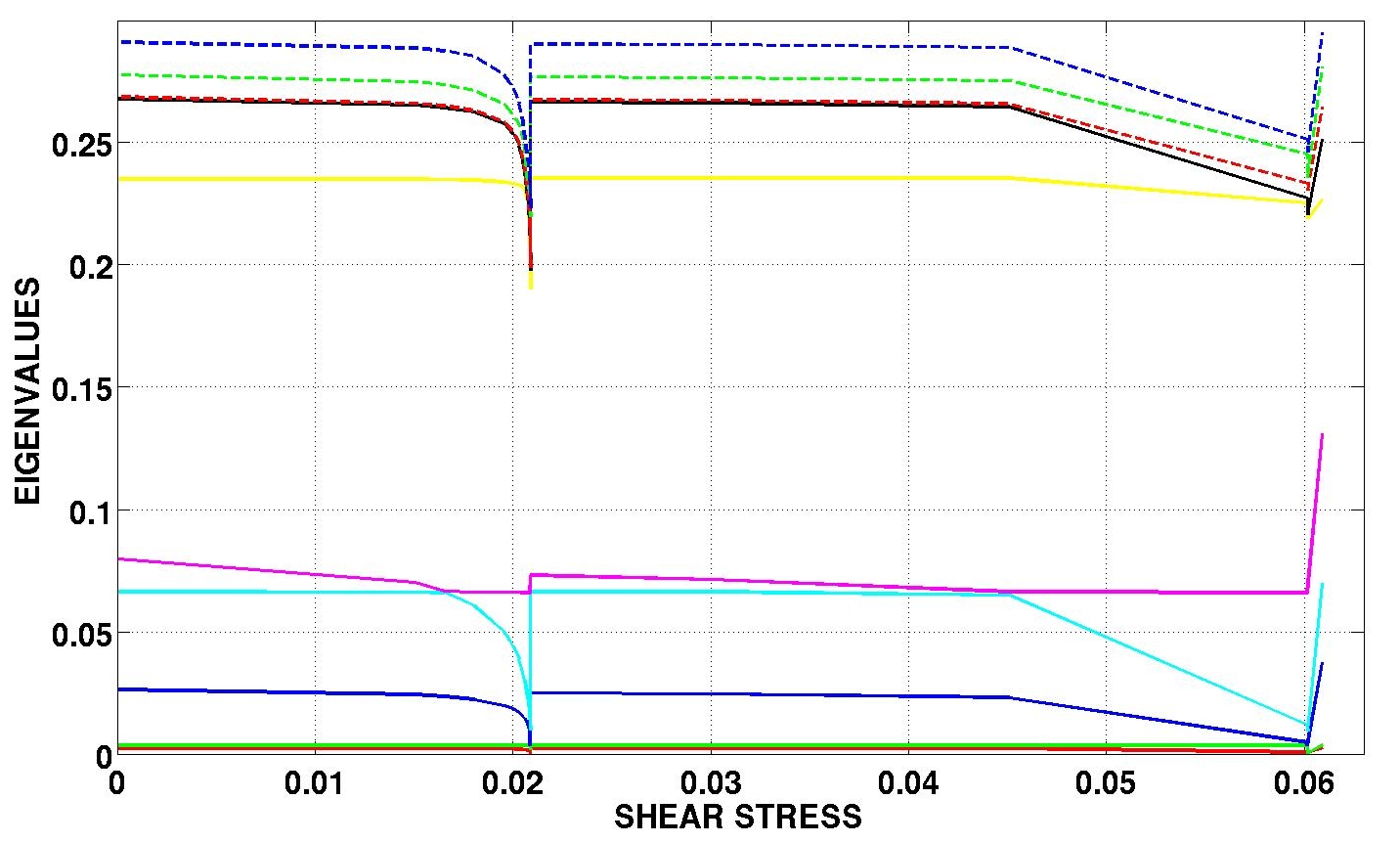}}
    \subfloat[Before the second drop.]
        {\label{fig:nx01ny06s02m03current036}\includegraphics[width=0.3\textwidth]{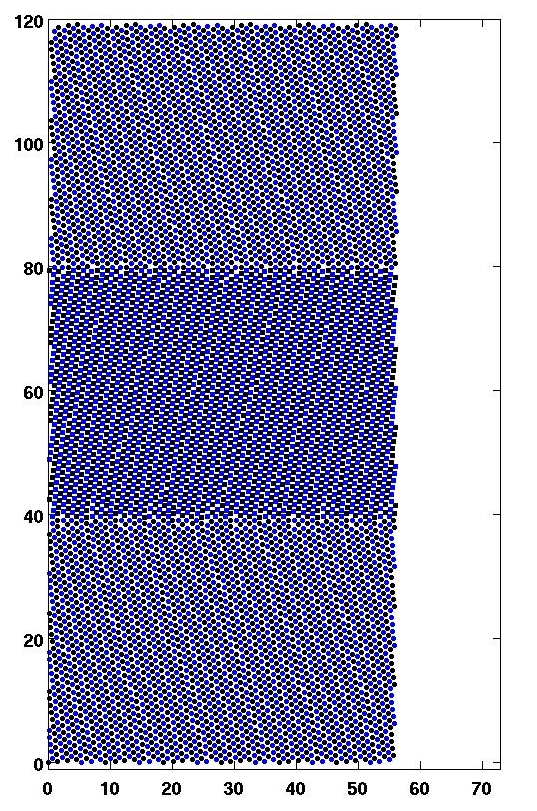}}
    \subfloat[After the second drop.]
        {\label{fig:nx01ny06s02m03current037}\includegraphics[width=0.3\textwidth]{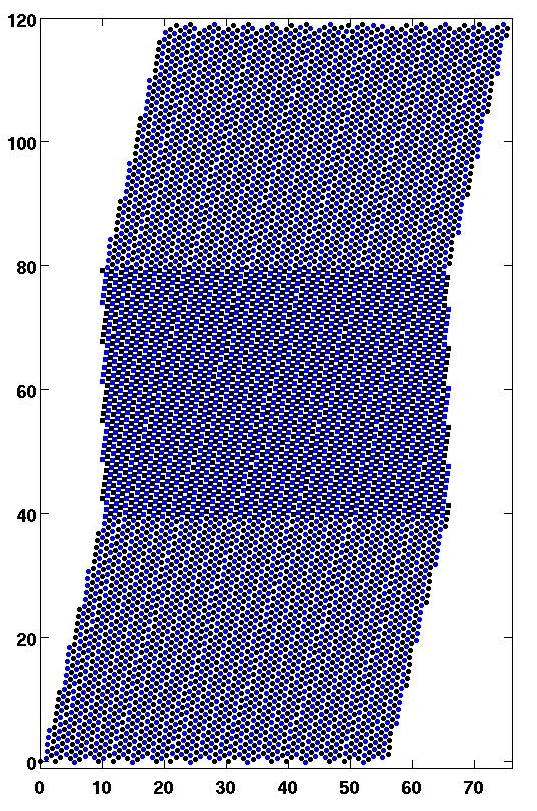}}
    \caption{Twin boundary motion for $\bfa=\left[160\right]_{2}$: (a) shows the evolution of the 10 lowest eigenvalues with load; (b) and (c) show the atomic configuration after the first and second load drops. The upper twin boundary moves upwards and the lower one moves downwards.}
    \label{fig:nx01ny06s02m03}
\end{figure}

\begin{figure}[htb!]
    \centering
    \mbox{}
    \hfill
    \subfloat[Eigenmode around the right edge of the upper twin boundary corresponding to the first drop in Fig. \ref{fig:nx01ny06s02m03eigenvalues}.]
        {\label{fig:nx01ny06s02m03mode026up}\includegraphics[width=0.35\textwidth]{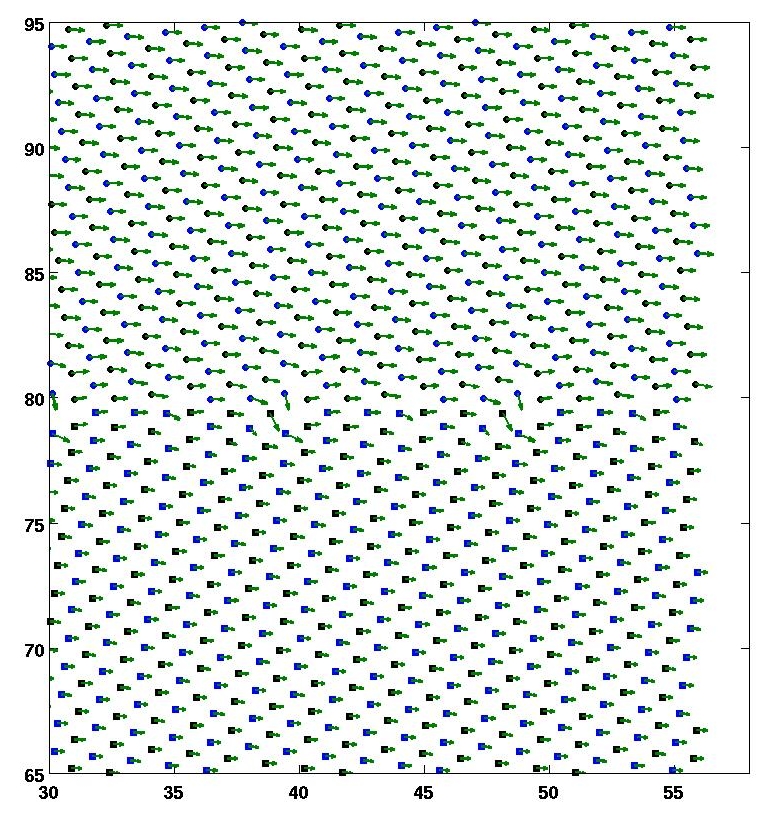}}
    \hfill
    \subfloat[Displacement around the right edge of the upper twin boundary when going through the first drop in Fig. \ref{fig:nx01ny06s02m03eigenvalues}.]
        {\label{fig:nx01ny06s02m03disp026up}\includegraphics[width=0.35\textwidth]{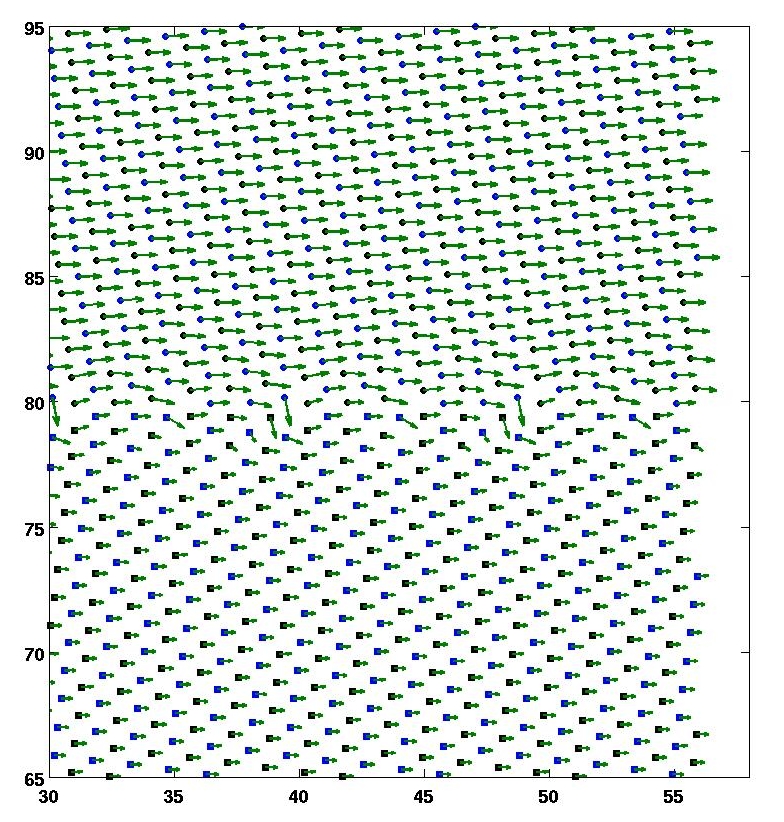}}
    \hfill
    \mbox{}
    \\
    \subfloat[Eigenmode around the upper twin boundary corresponding to Fig. \ref{fig:nx01ny06s02m03current036} (the second drop).]
        {\label{fig:nx01ny06s02m03mode036}\includegraphics[width=0.48\textwidth]{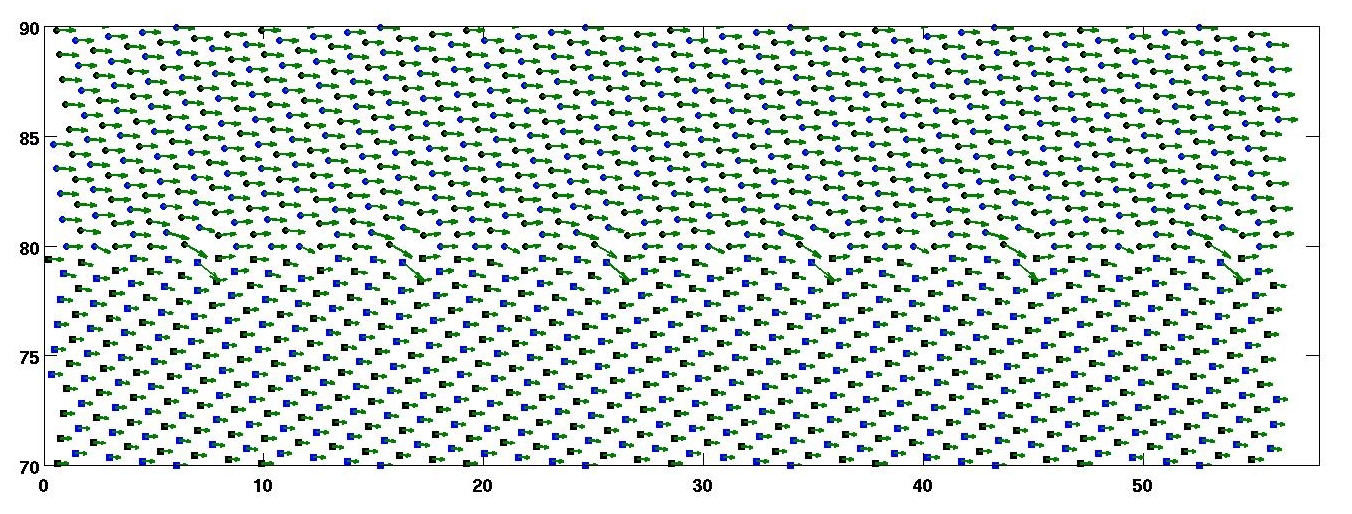}}
    \hfill
    \subfloat[Displacement around the upper twin boundary from Fig. \ref{fig:nx01ny06s02m03current036} and Fig. \ref{fig:nx01ny06s02m03current037} (going through the second drop).]
        {\label{fig:nx01ny06s02m03disp036}\includegraphics[width=0.48\textwidth]{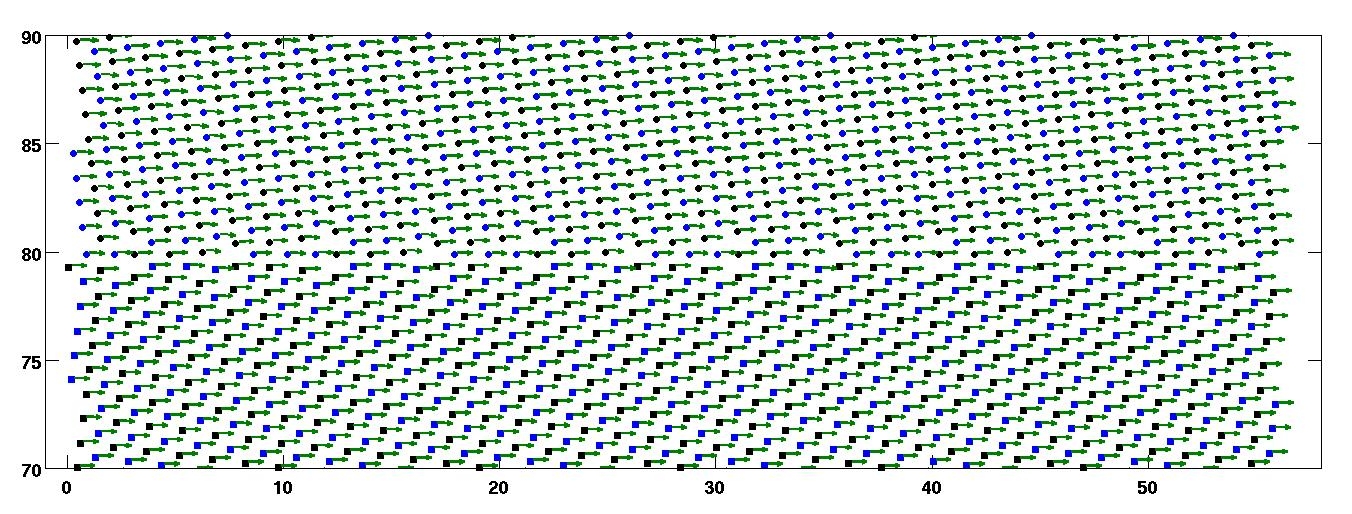}}
    \caption{Comparison of the soft eigenmodes and the observed displacements for $\bfa=\left[160\right]_{2}$ at the first eigenvalue drop, zoomed in to the vicinity of the twin interfaces. We see that the eigenmodes predict the displacements accurately, for both the first and second drops.}
    \label{fig:nx01ny06s02m03-eigenmode}
\end{figure}

We first consider for illustration the two solutions for $\bfa=\left[160\right]$.
These twin boundaries have a moderate level of rationality, i.e., the period to repeat along the interface is neither very large nor very small.
Figures \ref{fig:nx01ny06s01a01} -- \ref{fig:nx01ny06s02m03-eigenmode} show the evolution of the eigenvalues under load, the soft eigenmode when an eigenvalue goes to $0$, and the deformation of the lattice due to the motion of the twin interface.

We first consider $\bfa=\left[160\right]_{1}$, with the subscript $1$ denoting the first solution.
Figure \ref{fig:nx01ny06s01a01} shows the evolution of the eigenvalues with load: there are distinct values of the load at which the eigenvalues drop, with the lowest eigenvalue going to $0$.
These correspond precisely to the onset of twin boundary motion.
Further, Figure \ref{fig:nx01ny06s01a01-eigenmode} shows that the eigenmodes corresponding to the vanishing eigenvalue predict the observed displacements of the atoms well.

%%%%%%%%%%%%%%%%%%%%%%%%%%%%%%%%%%%%%%%%%%%%%%%%%

%%%%%%%%%%%%%%%%%%%%%%%%%%%%%%%%%%%%%%%%%%%%%%%%%

%%%%%%%%%%%%%%%%%%%%%%%%%%%%%%%%%%%%%%%%%%%%%%%%%

Next, we consider the interface $\bfa=\left[160\right]_{2}$.
Figure \ref{fig:nx01ny06s02m03} shows the evolution of the eigenvalues, and their drops precisely signal the initiation of motion.
Further, Figure \ref{fig:nx01ny06s02m03mode026up} clearly shows the soft eigenmode at the first drop, and it matches very well with the corresponding observed displacement in Figure \ref{fig:nx01ny06s02m03disp026up}.
However, at the second drop, the twin boundaries move and continue to propagate until they reach the top and bottom of the unit cell, i.e., the entire system is transformed to a  single twin variant.
Hence, the observed displacement in Figure \ref{fig:nx01ny06s02m03disp036} resembles a uniform translation.
The eigenmode in Figure \ref{fig:nx01ny06s02m03mode036} not only predicts this, but also provides more detail on how the motion is initiated.
Further, comparing Figure \ref{fig:nx01ny06s01a01eigenvalues} and Figure \ref{fig:nx01ny06s02m03eigenvalues}, we see that the critical shear stress that causes instability in $\bfa=\left[160\right]_{2}$ is significantly lower than that in $\bfa=\left[160\right]_{1}$.

%%%%%%%%%%%%%%%%%%%%%%%%%%%%%%%%%%%%%%%%%%%%%%%%%

%%%%%%%%%%%%%%%%%%%%%%%%%%%%%%%%%%%%%%%%%%%%%%%%%

\subsection{Significantly Lower Initiation Stress for Irrational Twins}\label{subsec:observe-1}

\begin{figure}[htb!]
  \centering
    \includegraphics[width=0.6\textwidth]{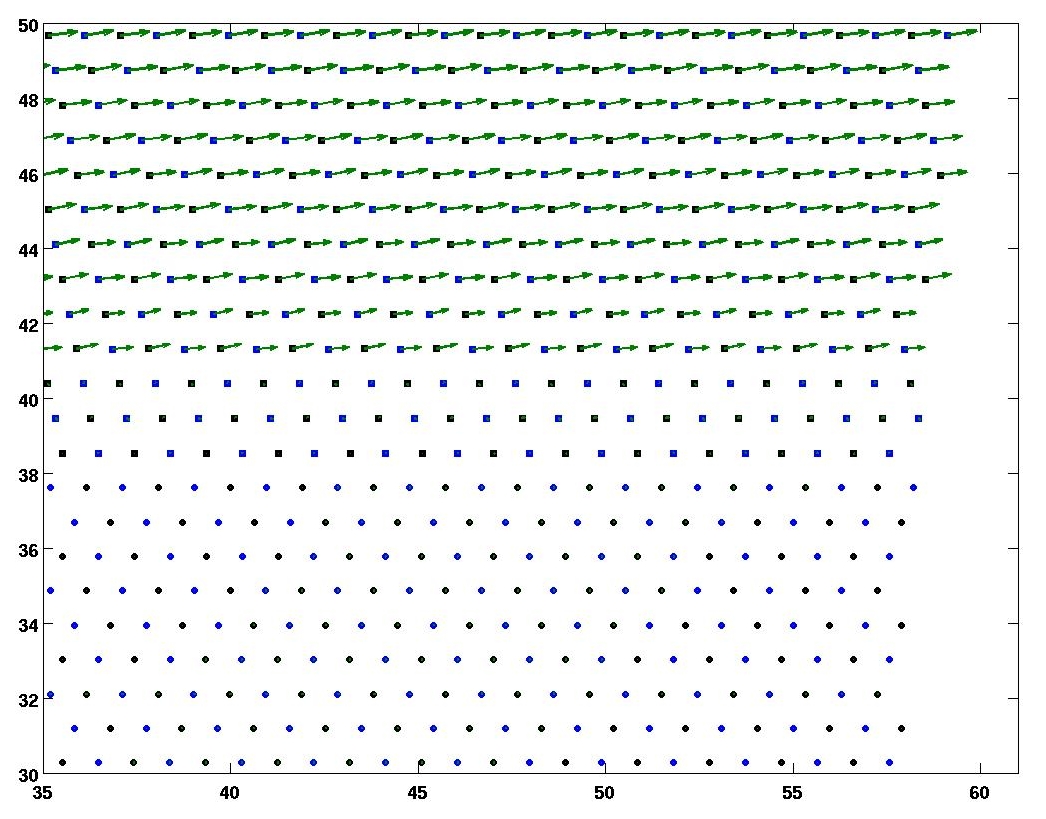}
  \caption{The soft eigenmode for $\bfa=\left[110\right]$.}
  \label{fig:nx01ny01-detwin}
\end{figure}

\begin{figure}[htb!]
  \centering
    \includegraphics[width=0.6\textwidth]{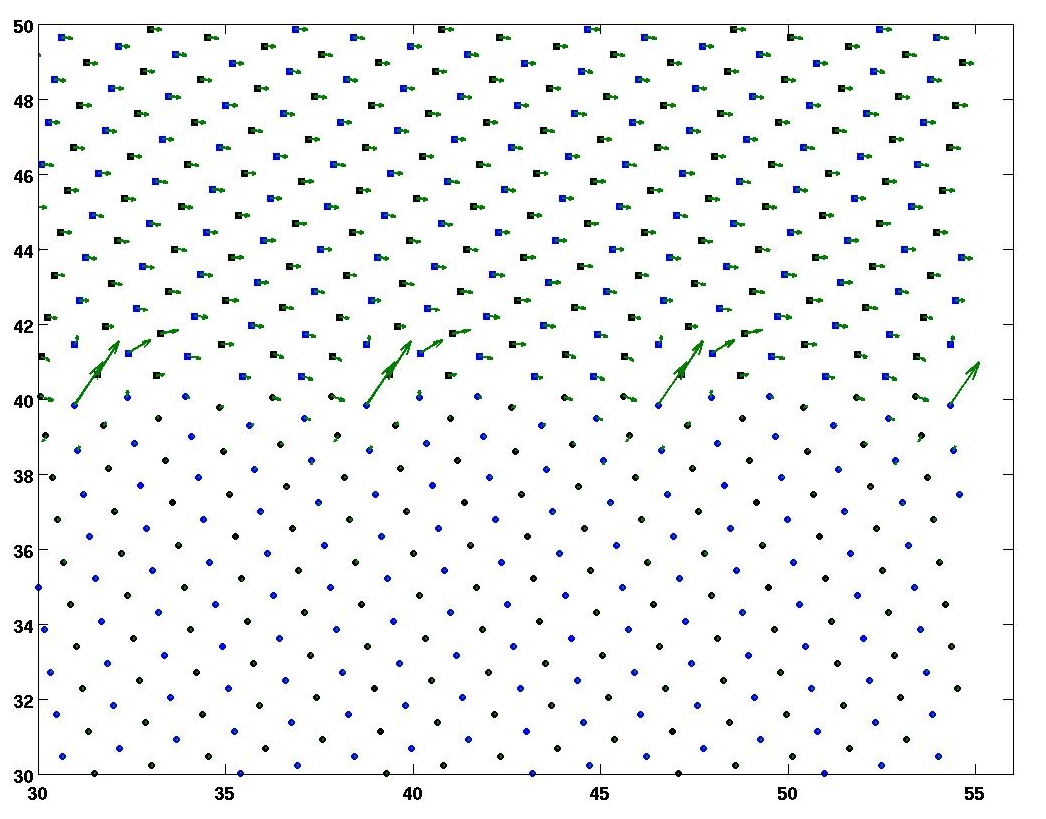}
  \caption{The soft eigenmode for $\bfa=\left[150\right]_{2}$.}
  \label{fig:nx01ny05s02a01-detwin}
\end{figure}

\begin{figure}[htb!]
  \centering
    \includegraphics[width=0.6\textwidth]{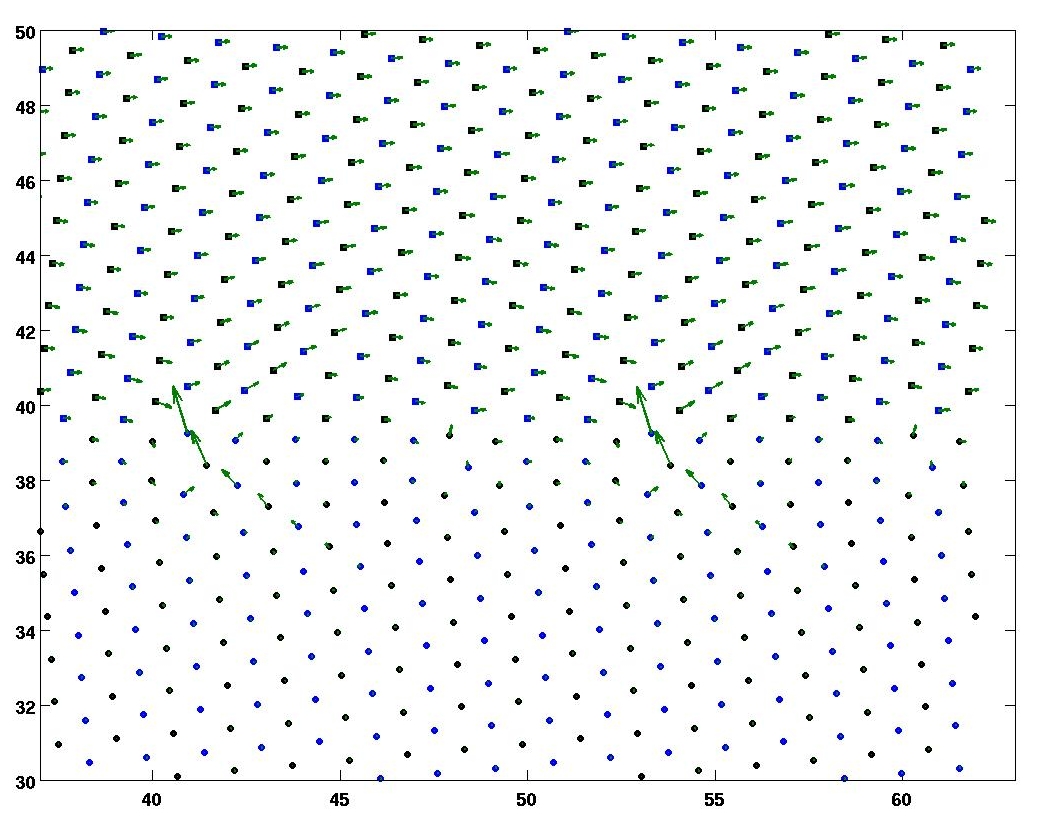}
  \caption{The soft eigenmode for $\bfa=\left[180\right]_{2}$.}
  \label{fig:nx01ny08s02m01-detwin}
\end{figure}

We consider the set of examples in Figure \ref{fig:nx01ny05s02a01-detwin} for $\bfa=\left[150\right]_{2}$, Figure \ref{fig:nx01ny06s02m03mode026up} for $\bfa=\left[160\right]_{2}$, and Figure \ref{fig:nx01ny08s02m01-detwin} for $\bfa=\left[180\right]_{2}$, where the soft eigenmode is shown.
These interfaces are much less rational than $\bfa=\left[110\right]$, whose soft eigenmode is shown in Figure \ref{fig:nx01ny01-detwin}.
However, a key common feature is that one of the lattice vectors is well-aligned with the plane of twin interface.
Further, in each of these cases, the large majority of atoms above the twin boundary move in synchrony along the shearing direction.

However, we also observe that several atoms close to the twin boundary initially move almost normal to the shearing direction.
This is in contrast to Figure \ref{fig:nx01ny01-detwin} for $\bfa=\left[110\right]$, where the motion of the atoms is completely uniform.
In the irrational interfaces, the bonding environment of these anomalous atoms is significantly different from that of the atoms in the bulk away from the interface.
While every atom in the bulk has 4 nearest neighbors that form a rectangular unit cell, the 4 atoms surrounding each of the anomalous atoms form an irregular quadrilateral or may not even have all 4 neighbors.
This provides a driving force and free volume to the anomalous atoms to move such that they are able to form the energetically-preferred regular rectangles by changing their configuration.
Further, the magnitudes of the displacements along the shearing direction are generally much smaller than the magnitudes of the displacements of those special atoms.

This mechanism is also reflected in the critical shear to initiate the motion, which is significantly lower for all the irrational twins.
In nondimensional terms, the critical shear stresses $\tau^{cr}$ are $\tau_{\left[110\right]}^{cr}=1.09136$ whereas $\tau_{\left[150\right]_{2}}^{cr}=0.23738$,$\tau_{\left[160\right]_{2}}^{cr}=0.193072$, and $\tau_{\left[180\right]_{2}}^{cr}=0.213072$.

%%%%%%%%%%%%%%%%%%%%%%%%%%%%%%%%%%%%%%%%%%%%%%%%%

%%%%%%%%%%%%%%%%%%%%%%%%%%%%%%%%%%%%%%%%%%%%%%%%

%%%%%%%%%%%%%%%%%%%%%%%%%%%%%%%%%%%%%%%%%%%%%%%%%

%%%%%%%%%%%%%%%%%%%%%%%%%%%%%%%%%%%%%%%%%%%%%%%%%

\subsection{Microtwinning Normal to the Primary Twin Interface}\label{subsec:observe-2}

\begin{figure}[htb!]
    \centering
    \subfloat[]
        {\label{fig:nx03ny04s01a01-5-detwin}\includegraphics[width=0.47\textwidth]{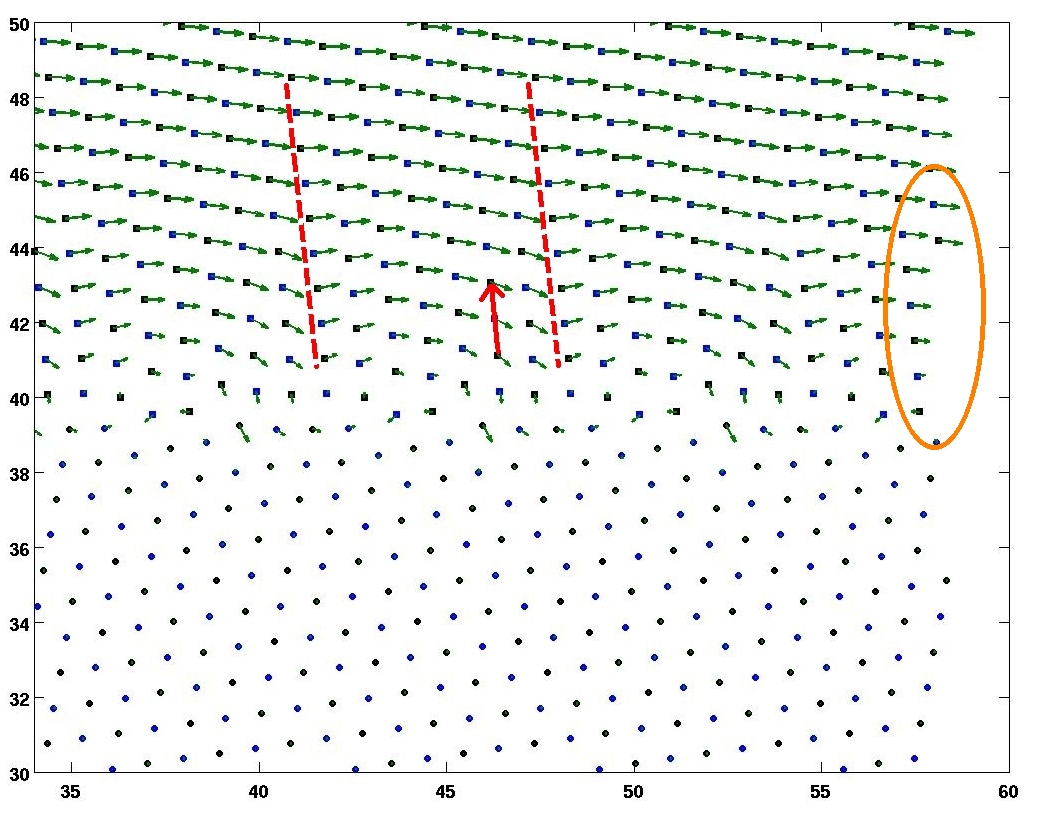}}
    \subfloat[]
        {\label{fig:nx03ny04s01a01-6-detwin}\includegraphics[width=0.47\textwidth]{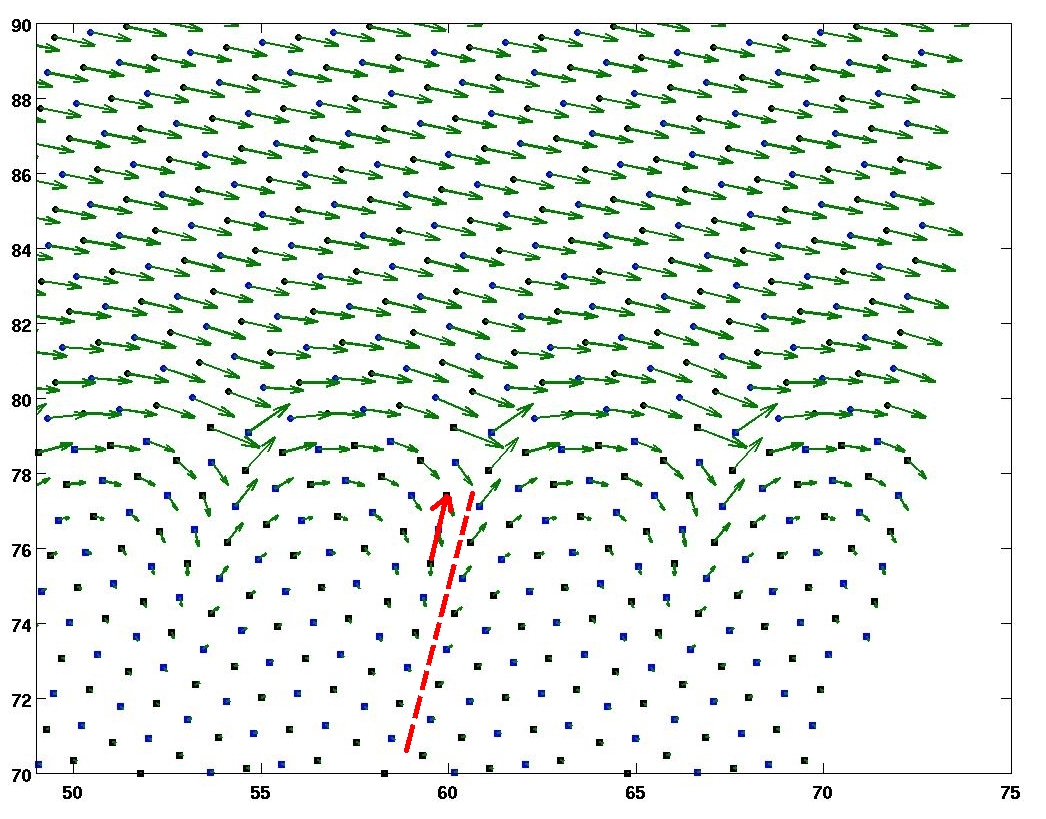}}
    \\
    \subfloat[]
        {\label{fig:nx04ny07s01a01-5-detwin}\includegraphics[width=0.47\textwidth]{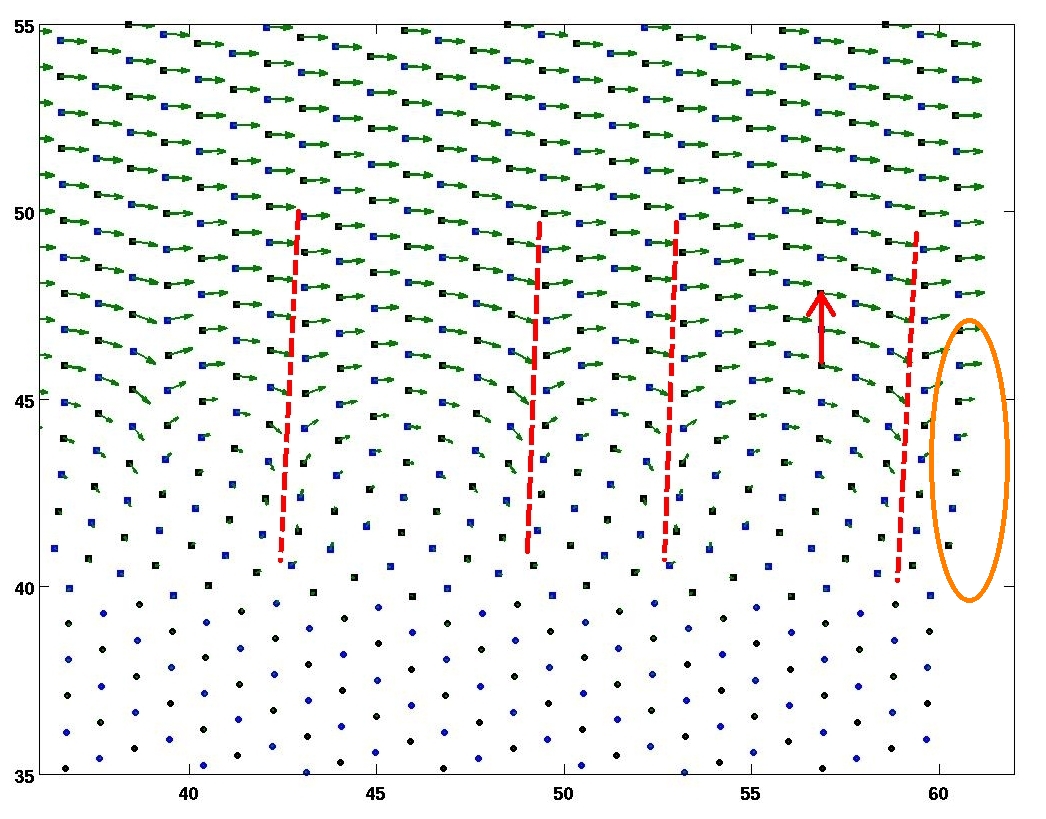}}
    \subfloat[]
        {\label{fig:nx04ny07s01a01-6-detwin}\includegraphics[width=0.47\textwidth]{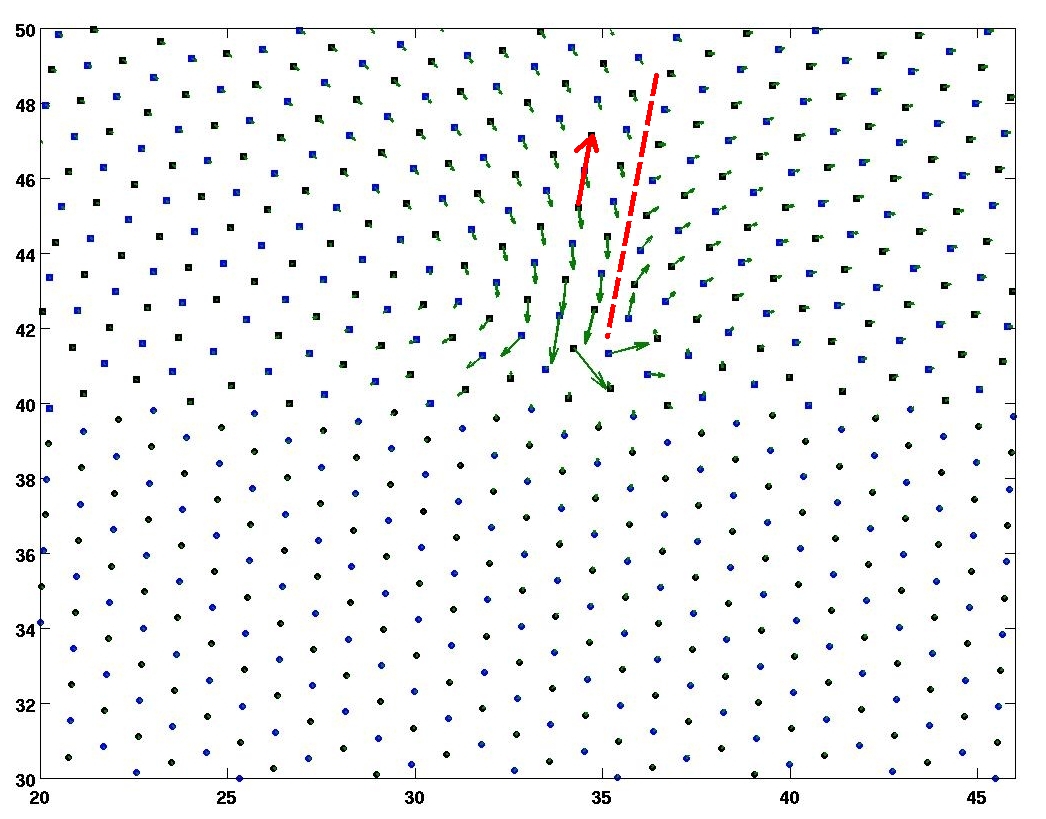}}
      \caption{Microtwin formation in $\bfa=\left[340\right]_{1}$ (top row) and $\bfa=\left[470\right]_{1}$ (bottom row). (a), (c) show the soft eigenmode with the nucleation of microtwins indicated by the red dashed lines in the diagonal direction along $\bff_{diag}$ shown by the red arrow. (b), (d) show that further loading causes the detwinning of the microtwins.}
    \label{fig:nx03ny04s01a01-detwin}
\end{figure}

We next consider the examples in Figure \ref{fig:nx03ny04s01a01-detwin} for $\bfa=\left[340\right]_{1}$ and $\bfa=\left[470\right]_{1}$.
A key common feature of these twin boundaries, as opposed to those considered above, is that the \textit{diagonal lattice vector} $\bff_{diag} := \half\left(\bff_1 + \bff_2\right)$ is almost normal to the twin boundaries.
When shear loads are applied, rather than causing the existing twin boundaries to move, the shearing results in the nucleation of new microtwin boundaries along $\bff_{diag}$ because this presents a lower energy deformation than twinning along the shearing direction.
Upon shearing further, detwinning occurs at the newly nucleated twin boundaries.
This is observed in many other cases that satisfy  $1<\frac{j}{i}<2$ in $\left[ij0\right]_{1}$, e.g., $\left[350\right]_{1}$ and $\left[450\right]_{1}$.

%%%%%%%%%%%%%%%%%%%%%%%%%%%%%%%%%%%%%%%%%%%%%%%%%

%%%%%%%%%%%%%%%%%%%%%%%%%%%%%%%%%%%%%%%%%%%%%%%%%

\begin{figure}[htb!]
    \centering
    \subfloat[]
        {\label{fig:nx05ny06s01a01-1-detwin}\includegraphics[width=0.47\textwidth]{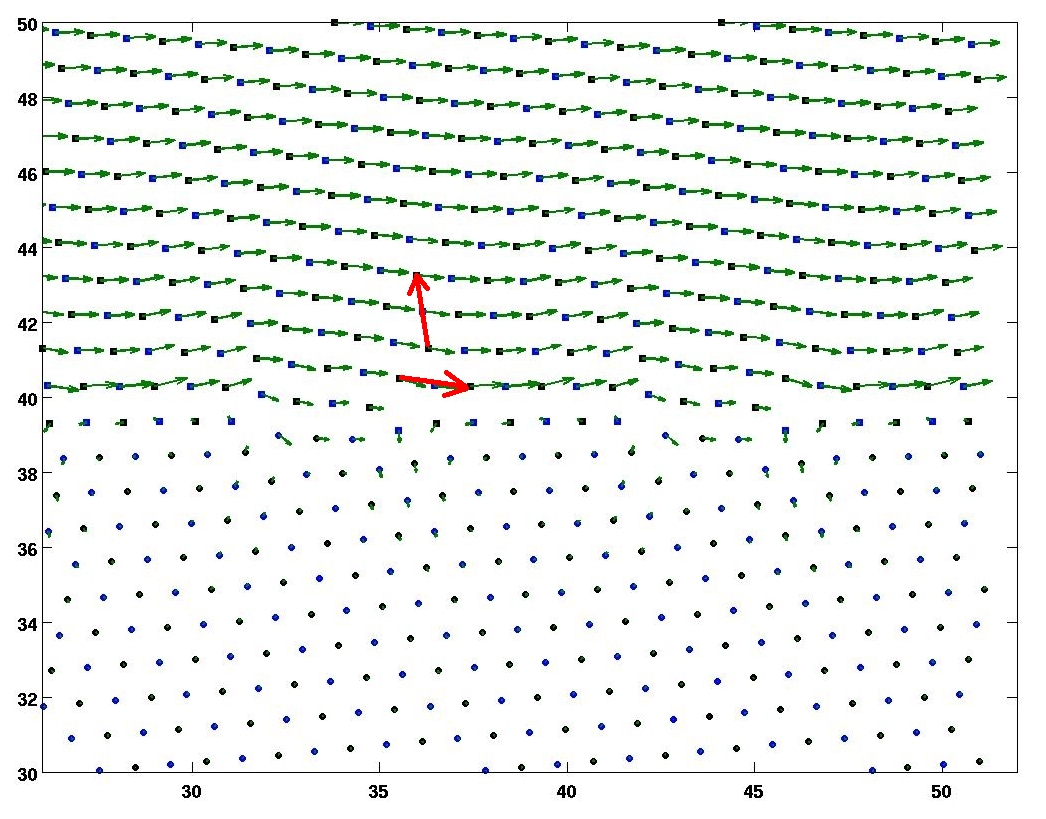}}
    \subfloat[]
        {\label{fig:nx05ny06s01a01-2-detwin}\includegraphics[width=0.47\textwidth]{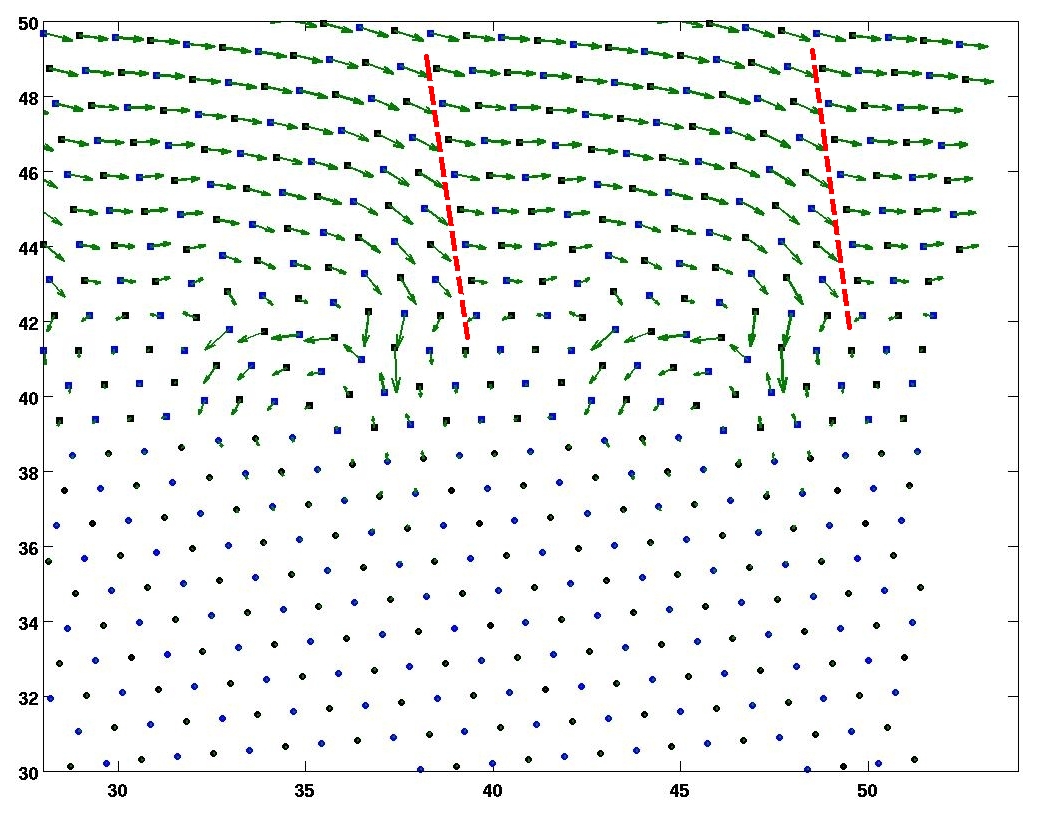}}
    \caption{Detwinning of $\bfa=\left[560\right]_{1}$ occurs by an alternation of (a) motion along the dominant shear direction, and then (b) the formation of microtwins along $\bff_{diag}$.}
    \label{fig:nx05ny06s01a01-detwin}
\end{figure}

Figure \ref{fig:nx05ny06s01a01-detwin} for $\left[560\right]_{1}$ is somewhat different, with $\frac{j}{i} \simeq 1$.
The zero eigenmodes first show that the detwinning occurs along the original twin boundaries when the shear is applied.
At a certain point after that, new twin boundaries are nucleated along $\bff_{diag}$ which is nearly normal to the twin boundaries.
The next detwinning event then takes place again along the original twin boundaries.
This suggests that the energetic cost for detwinning in either diagonal direction is comparable, and the direction of detwinning depends sensitively on the current deformation state.
The twin interfaces corresponding to $\left[570\right]_{1}$ and $\left[580\right]_{1}$ behave similarly.

%%%%%%%%%%%%%%%%%%%%%%%%%%%%%%%%%%%%%%%%%%%%%%%%%

%%%%%%%%%%%%%%%%%%%%%%%%%%%%%%%%%%%%%%%%%%%%%%%%%

\begin{figure}[htb!]
    \centering
    \subfloat[{$\bfa=\left[290\right]_{1}$}]
        {\label{fig:nx02ny09s01m01-slip}\includegraphics[width=0.47\textwidth]{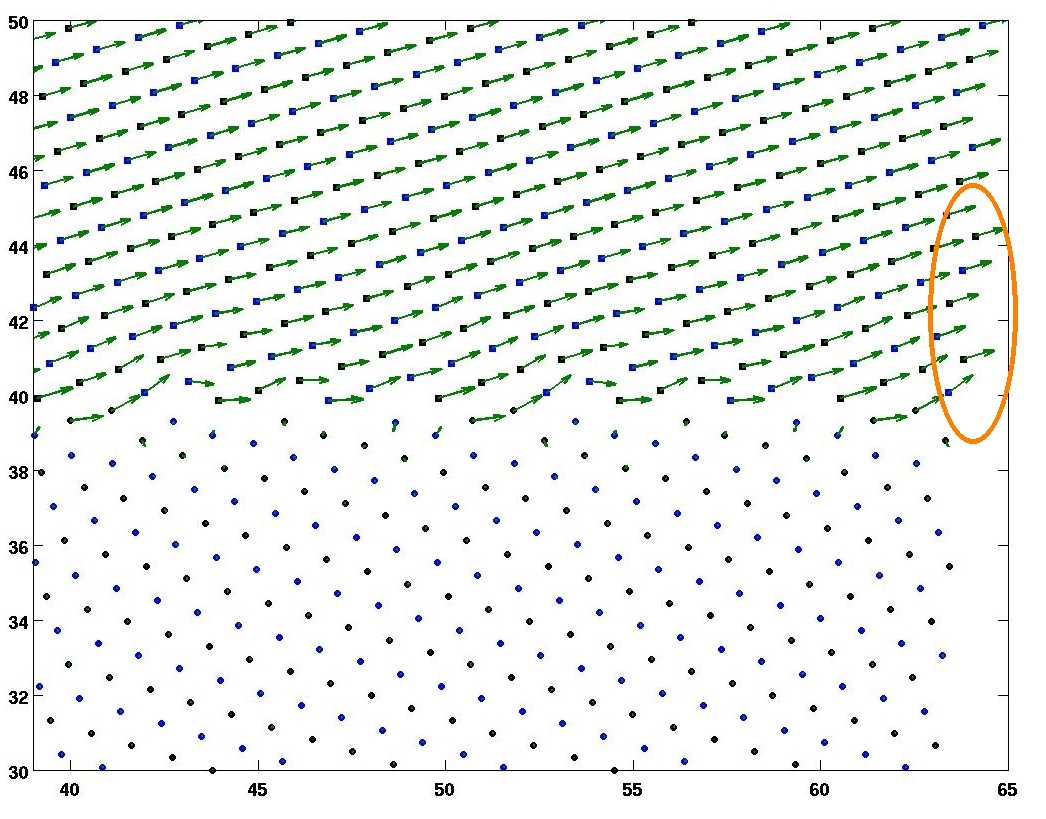}}
    \subfloat[{$\bfa=\left[350\right]_{1}$}]
        {\label{fig:nx03ny05s02a01-slip}\includegraphics[width=0.47\textwidth]{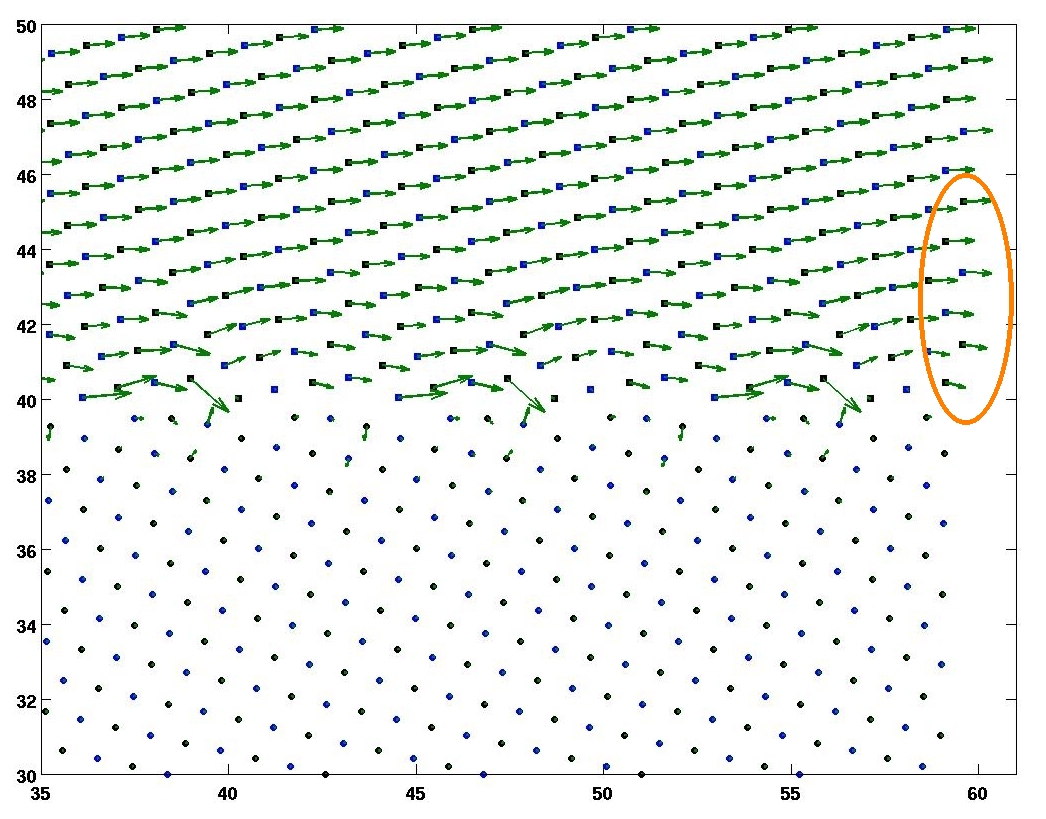}}
    \caption{Slip along the twin boundaries. The eigenmodes above the twin boundaries are so uniform that look like rigid body translation.}
    \label{fig:slipping}
\end{figure}

Finally, several other cases show deformations that appear as an almost rigid slip, with the atoms above the interface moving completely in sync (Fig. \ref{fig:slipping}).
However, the soft eigenmodes also indicate that very close to the twin boundaries, some atoms move slightly out of sync to avoid their neighbors.
Notice that the eigenmodes that predict this kind of motion look quite similar to the ones observed in \ref{subsec:observe-2} when detwinning occurs along $\bfF_{diag}$.
However, the magnitudes of the displacements near the twin boundaries are very uniform (highlighted by the orange ellipses), whereas the magnitudes of the displacements in \ref{subsec:observe-2} vanish with a gradient at the twin boundaries.

%%%%%%%%%%%%%%%%%%%%%%%%%%%%%%%%%%%%%%%%%%%%%%%%%
%%%%%%%%%%%%%%%%%%%%%%%%%%%%%%%%%%%%%%%%%%%%%%%%%
\subsection{Local Measures: Surface Atom Density, Surface Energy Density, and Maximum Energy per Atom}\label{subsec:atom-energy-emax}

The atoms around the twin boundaries have significantly different crystallographic environments compared to those that are away from the twin boundaries.
Therefore, we can think of various measures that capture these differences and then examine the correlation between boundary motion and these measures.
For instance, the atomic density and the energy per atom are different for atoms near the twin boundary; one could consider if, for example, a higher energy per atom would make the boundary easier to move?
From physical considerations \cite{sutton-balluffi}, we might expect that: (1) a higher atomic density on the interface makes motion easier because atoms need to move a shorter distance, although on the other hand, there is less available space for atoms to move; (2) a higher energy per atom on the interface suggests that atoms are more likely to move to find lower energy states; and (3) if there are some atoms with a higher energy per atom, these atoms are more likely to initiate motion.
We highlight that all of these are \textit{local} measures, as opposed to a linear stability analysis which is global.

The surface energy density for a twin boundary is calculated by computing the total energy of the system, then subtracting off the energy density of the bulk phases neglecting the interface, giving the excess energy due to the twin boundary. This excess energy is divided by the nominal area to find the surface energy density.
An analogous approach gives the atomic density on the interface, by replacing energy by number of atoms.
Finally, we can compute the highest energy per atom by simply identifying the atom in each case with the highest energy.

\begin{figure}[htb!]
    \centering
    \subfloat[]
        {\label{fig:atom_density}\includegraphics[width=0.33\textwidth]{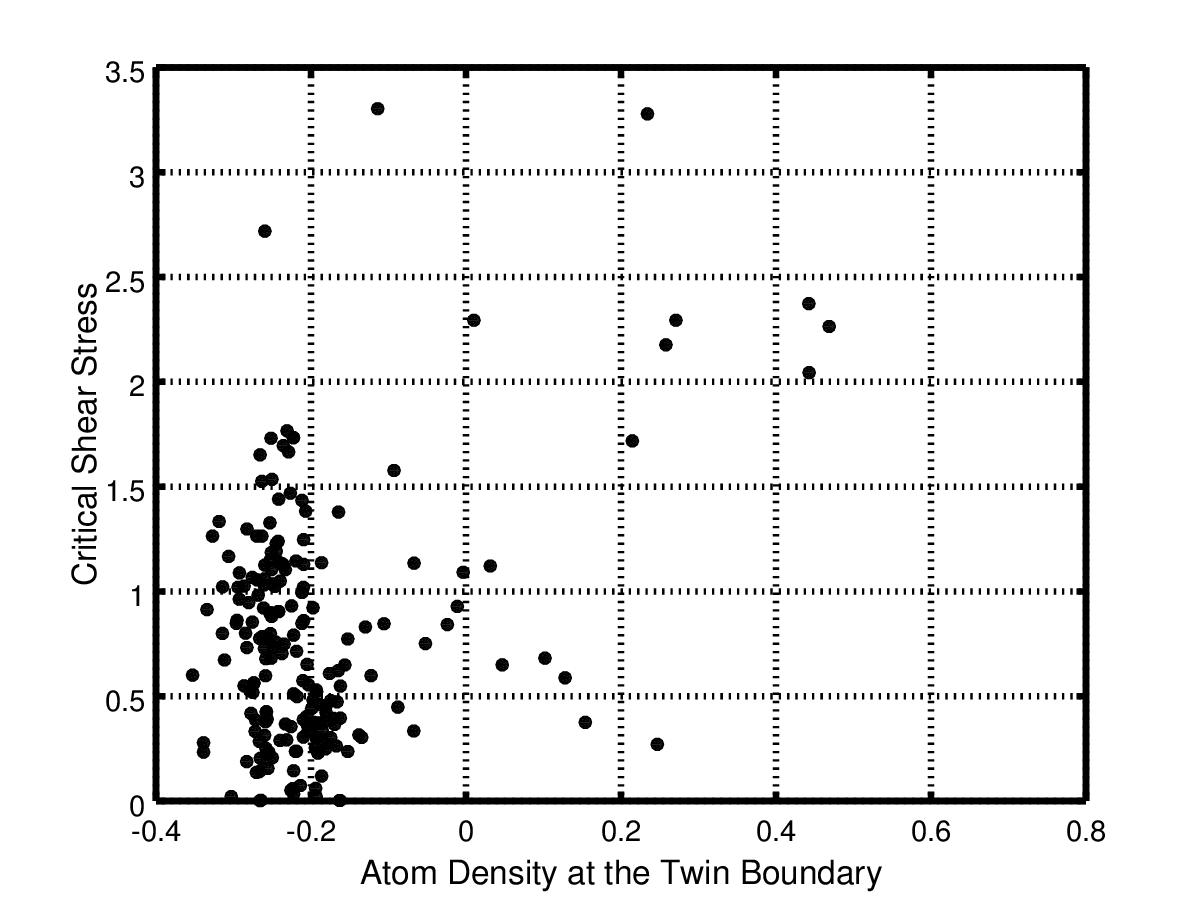}}
    \subfloat[]
        {\label{fig:energy_density}\includegraphics[width=0.33\textwidth]{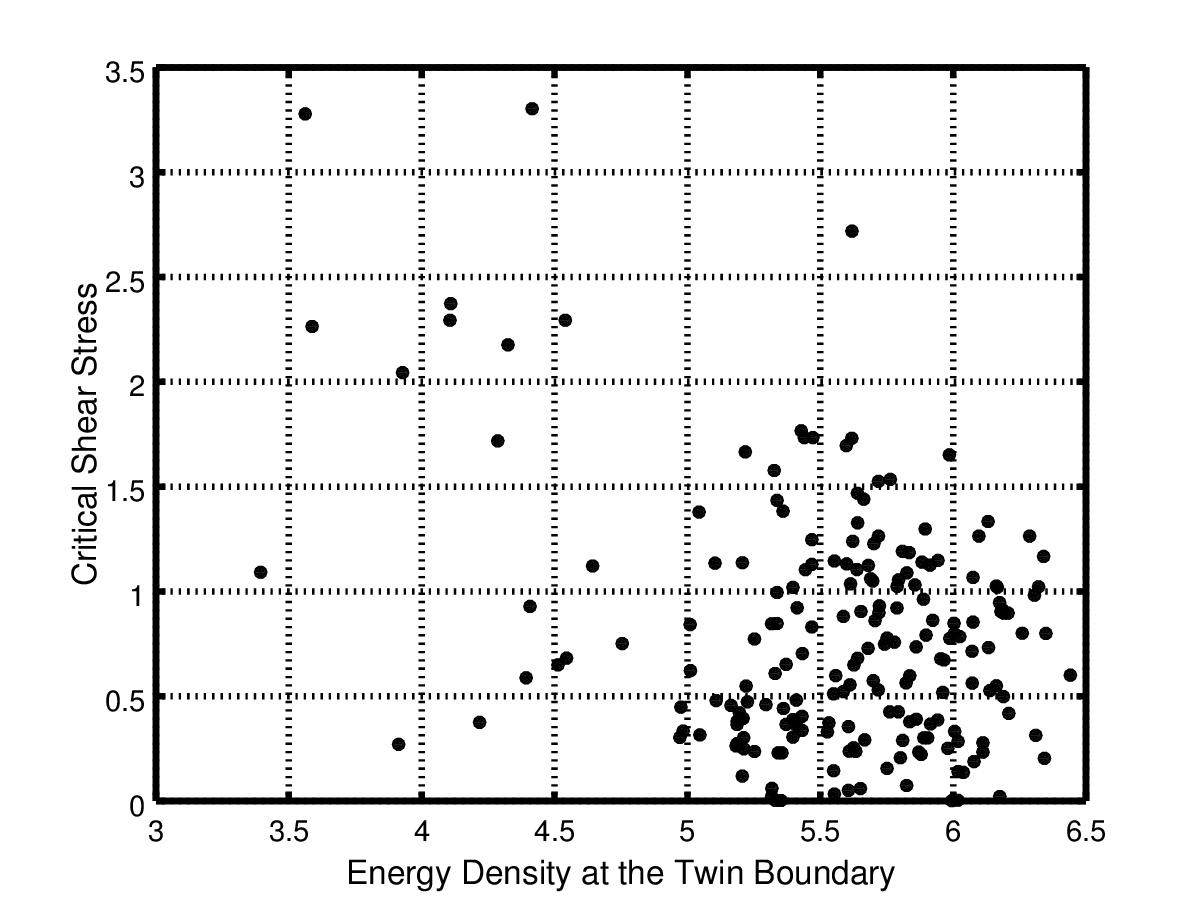}}
    \subfloat[]
        {\label{fig:max_energy}\includegraphics[width=0.33\textwidth]{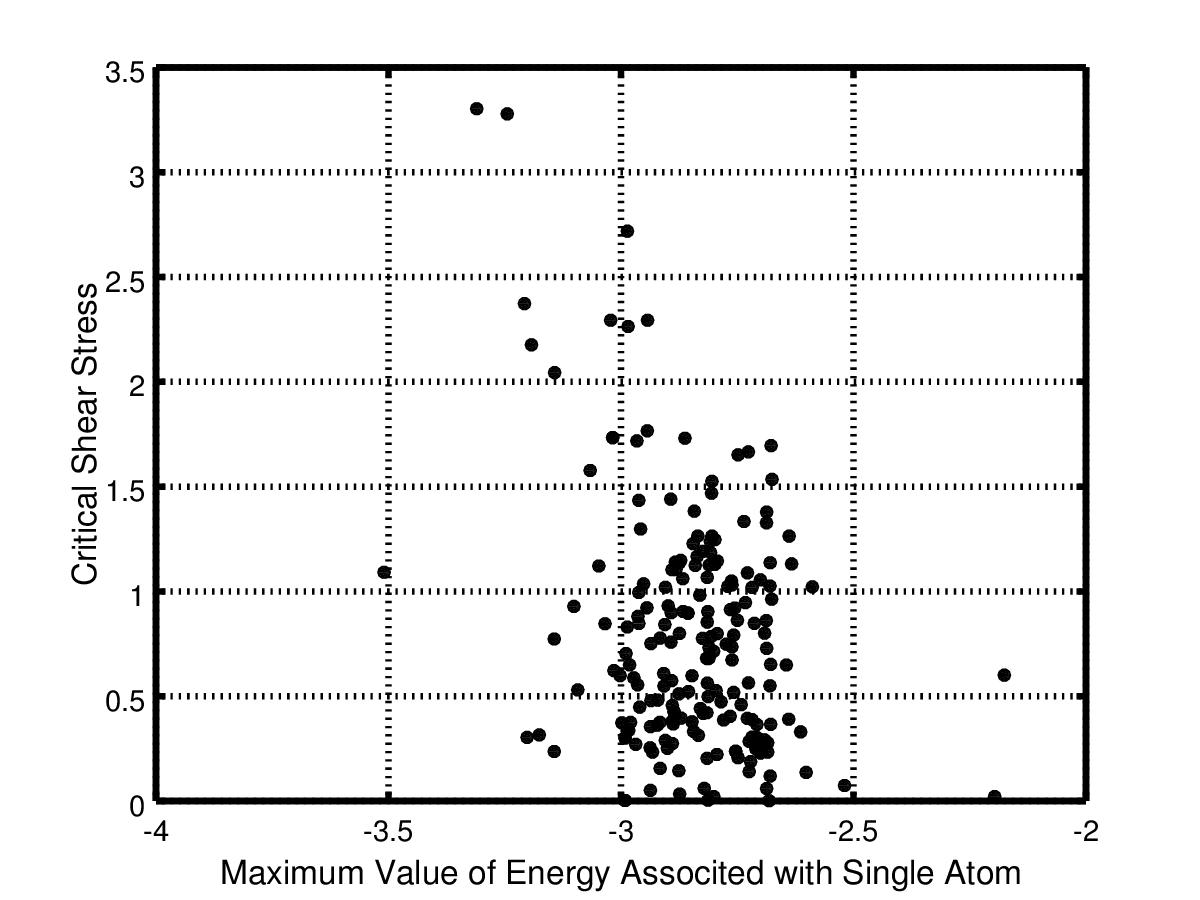}}
    \caption{Correlations for a large number of different twins between the critical shear stress for the initiation of motion between (a) the surface atomic density, (b) the surface energy density, and (c) the highest energy per atom, with all of these quantities calculated in the configuration before the motion initiates. We see that none these show strong correlations with the critical shear stress.}
    \label{fig:atom-energy-emax}
\end{figure}

Figure \ref{fig:atom-energy-emax} shows the correlations between all these quantities against the critical shear stress for the initiation of motion.
There are no clear correlations with any of these local measures, whereas the nonlocal linear stability analysis provided an accurate indicator of the initiation of motion.
This aligns with prior work on the nucleation and motion of defects of various types \cite{miller2008nonlocal,lu-dayal-2011,dayal-bhattacharya-2006}.

%%%%%%%%%%%%%%%%%%%%%%%%%%%%%%%%%%%%%%%%%%%%%%%%%

%%%%%%%%%%%%%%%%%%%%%
%%%%%%%%%%%%%%%%%%%%%
%%%%%%%%%%%%%%%%%%%%%
%%%%%%%%%%%%%%%%%%%%%
\section{Conclusions}

We have used molecular statics to investigate the initiation of motion of both rational and irrational twin boundaries under loading. 
We have shown that the onset of motion is governed by a linear instability, manifested by the vanishing of the lowest eigenvalue with the corresponding eigenmode accurately predicting the initial atomic displacements, showing the collective nature of the instability.
Comparisons with local energetic measures, such as interface energy density or number density, show no clear correlation with the onset of motion, reinforcing that the instability is a nonlocal phenomenon.

We apply linear stability analysis to show that irrational twins exhibit markedly lower critical shear stresses for motion initiation compared to rational twins. 
The associated eigenmodes for irrational twins display complex structure such as transverse microtwinning along the interface, suggesting that the non-periodic atomic environments at irrational interfaces drive heterogeneous rearrangements that enhance mobility. 
In contrast, rational interfaces exhibit higher initiation loads and more uniform instability modes, consistent with their periodic structure.
These insights bridge atomistic stability analysis and macroscopic kinetic behavior, providing a foundation for predictive, multiscale models of twin-mediated deformation in complex crystalline materials.

There are several open questions for the future.
A key shortcoming of linear stability is that it is strictly valid only in the zero-temperature limit; it is important to understand how finite-temperature thermal activation influences the onset of motion and kinetics. 
It is also important to go from these atomistic insights to develop continuum descriptions that embed the atomistically-derived instability criterion within mesoscale kinetic laws \cite{abeyaratne-knowles-book-2006}, to enable the prediction of twin motion in general loading conditions and geometries.

%%%%%%%%%%%%%%%%%%%%%
%%%%%%%%%%%%%%%%%%%%%
%%%%%%%%%%%%%%%%%%%%%
%%%%%%%%%%%%%%%%%%%%%

\begin{acknowledgments}
    We acknowledge financial support from ARO (MURI W911NF-24-2-0184) and NSF (DMREF 2118945); NSF for XSEDE computing resources provided by Pittsburgh Supercomputing Center; and Rohan Abeyaratne, Gregory Rohrer, and Robert Sekerka for useful discussions.
\end{acknowledgments}

% References

\end{document}